\documentclass[useAMS,usenatbib]{mn2e}
\usepackage[usenames,dvipsnames]{color}

\newcommand{\iso}[2] {$^{#1}{\rm #2}$}		
\newcommand{\Mm}{\ensuremath{\times10^8~{\rm cm}}} 
\newcommand{\msun}{\ensuremath{M_\odot}}
\newcommand{\lsun}{\ensuremath{L_\odot}}
\newcommand{\diff}{\mathrm{d}}
\newcommand{\rlb}{\ensuremath{r_\mathrm{lb}}}
\newcommand{\rub}{\ensuremath{r_\mathrm{ub}}}

\usepackage{graphicx}
\usepackage{amssymb}
\usepackage{amsmath}
\usepackage{color}
\usepackage{soul}
\usepackage{booktabs}
\usepackage{afterpage}
\usepackage{url}
\usepackage{enumerate}

\title[Idealised $4\pi$ simulations of O-shell convection]{Idealised
hydrodynamic simulations of turbulent oxygen-burning shell convection in $4\pi$
geometry}

\author[S.~W.~Jones et.~al]{
S.~Jones$^{1,2}$\thanks{E-mail: samuel.jones@h-its.org},
R.~Andrassy$^{2,3}$,
S.~Sandalski$^{4,3}$,
A.~Davis$^2$,
P.~Woodward$^{4,3}$  
\newauthor 
and F.~Herwig$^{2,3}$ \\
$^{1}$Heidelberg Institute for Theoretical Studies, Schloss-Wolfsbrunnenweg 35,
D-69118 Heidelberg, Germany\\
$^{2}$Department of Physics \& Astronomy, University of Victoria, P.O. Bos 3055
Victoria, B.C., V8W 3P6, Canada\\
$^{3}$Joint Institute for Nuclear Astrophysics, Center for the Evolution of the
Elements, Michigan State University, \\ 640 South Shaw Lane, East Lansing, MI
48824, USA\\
$^{4}$LCSE and Department of Astronomy, University of Minnesota, Minneapolis,
MN 55455, USA
}

\begin{document}

\date{Accepted 2016 October 27. Received 2016 October 25; in original form 2016
May 9}


\maketitle

\label{firstpage}

\begin{abstract} This work investigates the properties of convection in stars
	with particular emphasis on entrainment across the upper convective
	boundary (CB). Idealised simulations of turbulent convection in the
	O-burning shell of a massive star are performed in $4\pi$ geometry on
	$768^3$ and $1536^3$ grids, driven by a representative heating rate. A
	heating series is also performed on the $768^3$ grid. The $1536^3$
	simulation exhibits an entrainment rate at the upper CB of
	$1.33\times10^{-6}~\msun~\mathrm{s}^{-1}$. The $768^3$ simulation with
	the same heating rate agrees within 17 per cent. The entrainment rate
	at the upper convective boundary is found to scale linearly with the
	driving luminosity and with the cube of the shear velocity at the upper
	boundary, while the radial RMS fluid velocity scales with the cube root
	of the driving luminosity, as expected. The mixing is analysed in a 1D
	diffusion framework, resulting in a simple model for CB mixing. The
	analysis confirms previous findings that limiting the MLT mixing length
	to the distance to the CB in 1D simulations better represents the
	spherically-averaged radial velocity profiles from the 3D simulations
	and provides an improved determination of the reference diffusion
	coefficient $D_0$ for the exponential diffusion CB mixing model in 1D.
	From the 3D simulation data we adopt as the convective boundary the
	location of the maximum gradient in the horizontal velocity component
	which has $2\sigma$ spatial fluctuations of $\approx0.17 H_P$ . The
	exponentially decaying diffusion CB mixing model with $f = 0.03$
	reproduces the spherically-averaged 3D abundance profiles.
\end{abstract}


\section{Introduction}
Modelling the long-term evolution of stars requires the assumption of spherical
symmetry and hydrostatic equilibrium. Such models are crucial for predicting
the characteristics of stars with a given initial mass, metallicity,
composition and rotation rate, which are important for many areas of astronomy
(e.g.~distance, age and mass determinations, population synthesis, galactic
chemical evolution, etc.).  The predictive power of these models is somewhat
restricted by their dependence on approximate and often parametrised treatment
of underlying physical processes that become necessary when enforcing
symmetries and equilibria upon the models.  Two such processes are convective
mixing and the mixing at convective boundaries, for which either an
instantaneous mixing or diffusion approximation is typically used
\citep[e.g.][]{freytag:96, herwig:97, EldridgeTout2004, Young2005,
Eggenberger2008a, Brott2011, Limongi2012, Ekstrom2012a, Rauscher2002, WH07,
Paxton:2013km, MESA2015}. A key quantity of interest is the rate at which
material is entrained into convection zones, that is the rate at which material
is transported across the convective boundaries from neighbouring stable layers
by hydrodynamic instabilities in the vicinity of the convective boundary.

The varied ways in which these approximations are formulated and then
implemented into stellar evolution codes result in different codes making
different predictions for, e.g., nuclear burning lifetimes, core masses,
evolution in the Hertzsprung--Russell diagram, weak $s$-process production
\citep[e.g.][]{Martins2013, Jones:2015cc}. In the advanced burning stages of
massive stars (C-burning onwards), the structure of the star is generally said
to be frozen-in. However, as \citet{Sukhbold2014} have shown, seemingly
insignificant changes in the masses of convective regions within the stellar
core during the advanced burning stages can have a marked effect on the
interior core structure. These changes to the core structure can shift the
compact remnant that is predicted to be produced when the star dies, from a
neutron star to a black hole \citep{Ertl2016}. The former case would
typically result in a core-collapse supernova of type II\footnote{however, note
that note that a subset of type I supernovae (Ib and Ic) are also core-collapse
supernovae}, while the latter case (black hole formation) is predicted to
produce a long-duration ($\sim1$ year) low luminosity ($\sim
10^{39}$~erg~s$^{-1}$) transient often referred to as a \emph{failed} supernova
\citep{Lovegrove2013a}. To date, the only progenitor stars of type II
supernovae that have been detected fall in the luminosity range
$L\lesssim10^{5.1}~\lsun$, corresponding to an initial progenitor mass of
$M\lesssim18~\msun$ \citep{Smartt2015a}. However, red supergiant stars have
indeed been observed with luminosities in the range $10^{5.1}\lesssim
L/\lsun\lesssim 10^{5.5}$ \citep{Levesque2006a,Levesque2009a} and from a
statistical point of view based on the current theory of stellar evolution,
stars this luminous/massive should have contributed to the number of direct
detections mentioned previously. This suggests that red supergiants with
$L\gtrsim 10^{5.5}~\lsun$ ($M\gtrsim18~\msun$) always result in weak or failed
supernovae that form black holes. This is a picture that is not, however,
corroborated by recent spherically symmetric supernova simulation efforts
\citep{OConnor2011a,Ugliano2012a,Ertl2016,Sukhbold2016a} although the explosion
mechanism of core-collapse supernovae is still not completely understood, and
is certainly strongly influenced by asymmetries in the structure of the
progenitor star and in the explosion itself \citep[see][for a recent
review]{Mueller2016a}. With the outcome of such simulated explosions being so
intrinsically linked to the structure of the progenitor star, the accuracy of
stellar models is arguably as important as that of the supernova simulations
when distinguishing between a failed (black hole-forming) or successful
(neutron star-forming) supernova explosion for a star of given initial mass,
metallicity and rotation rate. The mass of the compact remnant produced during
core collapse is also strongly related to the pre-supernova stellar structure,
and different predictions for pre-SN structures as a function of initial
stellar mass and metallicity can lead to rather different conclusions. A timely
example is the metallicity upper limit placed on the progenitor system of the
binary black hole merger whose detection was the first of its kind using
gravitational wave telescopes \citep{Abbott2016a}. The binary population
synthesis code StarTrack placed a limit of $Z<0.1~Z_\odot$
\citep{Belczynski2016a} while the limit from another such code BPASS was
$Z<0.5~Z_\odot$ \citep{Eldridge2016a}, in agreement with the prediction of
\citet{Abbott2016b} which was obtained independently. Binary population
synthesis models, however, due to their very nature, inherit the significant
uncertainties of stellar models, and so to better constrain stellar models is
also to improve the predictions of population synthesis models, through which
the stellar models may be validated in the future with further gravitational
wave measurements from compact binary mergers.

\begin{figure}
\includegraphics[width=\linewidth, clip=true, trim=0mm 2mm 0mm 
                 3mm]{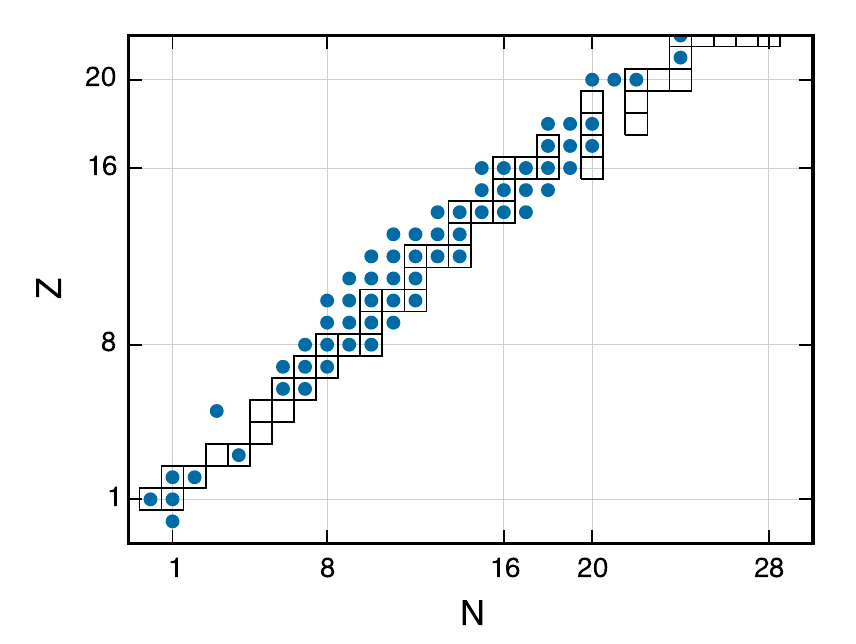}

\caption{Isotopes included in the nuclear reaction network in the
25~\msun~MESA model during oxygen shell burning (circles). Stable
isotopes are outlined with black squares.}

\label{fig:MESAnetwork}
\end{figure}

The structure of the core during the advanced burning stages of massive stars
is determined by many different aspects, which are themselves products of other
physical processes that have in some cases been described in a stellar model by
their own parametrisation. For example, the $^{12}$C/$^{16}$O ratio in the core
at the onset of C burning is a major factor setting the mass of the convective
C-burning core and hence the location of the first subsequent C-burning shell.
This ratio is set by the competition between the triple-alpha (3$\alpha$) and
the $^{12}$C($\alpha,\gamma)^{16}$O reactions at the end of the core He-burning
phase, which are sensitive to both the number of $\alpha$-particles and the
temperature. The $\alpha$-particles are introduced into the core at different
rates depending upon the treatment of mixing at the boundary of the convective
He-burning core, and the temperature is set by the mass of the H-depleted core,
which is itself determined by the treatment of convective boundary mixing
(CBM) during the main sequence (core H-burning). So, the core structure
of a massive star during O shell burning, for example, is not only determined
by the mixing assumptions in the O shell but in every convective episode prior.

Asteroseismological measurements of mixed modes in core He-burning stars
provide a constraint for the extent and behaviour of convective
(boundary/overshoot) mixing in stellar models \citep[e.g.][]{Constantino2015,
Bossini2015}. However, it is rather optimistic to expect that the same
diagnostics will be measured for an O-burning shell in a 25\msun~star.  What
can be done, however, are multidimensional hydrodynamic simulations of the
O-burning shell that describe the convective properties very well on short
timescales \citep{Arnett1994, Bazan1998, Asida2000, Young2005, Meakin:2007dj}.
These kinds of simulations are much more difficult during core He-burning
because of the much lower Mach number of the flow and the much longer
evolutionary time scales.

Mixing processes at the convective boundaries of advanced (C-, O- and Si-)
burning shells that can be studied in multidimensional hydrodynamic simulations
\citep[see, e.g.,][]{Arnett1994,Bazan1998,Asida2000,Young2005, Meakin:2007dj}
are known to have an impact on the pre-supernova structure of core-collapse
supernova (CCSN) progenitor models \citep{Young2005,Frey2013,Sukhbold2014},
which in turn affects the dynamics of the supernova explosion via an alteration
of the competition between the ram pressure of the in-falling core material
with the neutrino-heated material behind the stalled shock.  Because the time
scale of the collapse is much shorter than the time scale of convection in the
final O shell-burning episode, the imprint of the convective velocity field is
essentially frozen in.  Fluctuations in the velocity amplitudes can impact the
success of the simulated core-collapse supernova explosion of the progenitor
star \citep{Couch2013, Mueller2015, Couch2015}.

\begin{figure}

\includegraphics[width=\linewidth, clip=true, trim=0mm 0mm 0mm 
                 0mm]{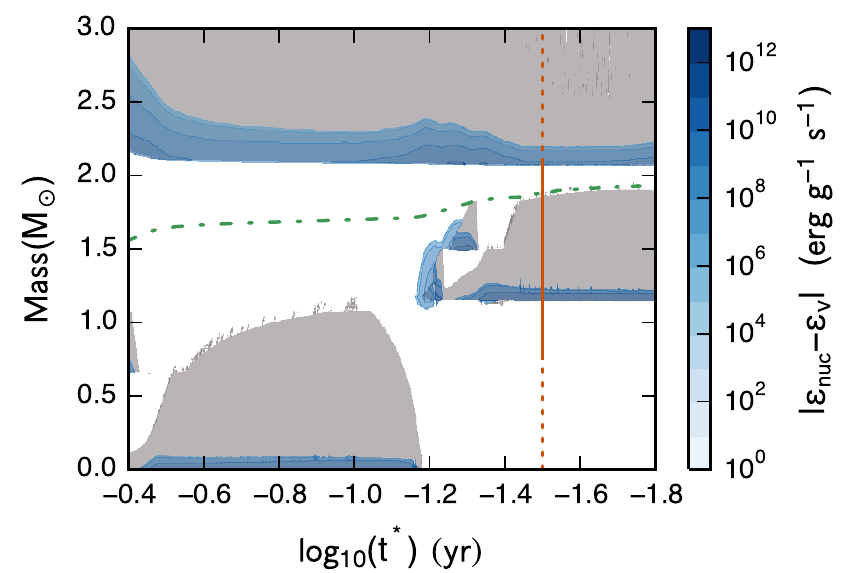}

\caption{Kippenhahn (convective structure evolution) diagram of the core and
shell oxygen-burning phases in the 25~\msun~MESA model with positive nuclear
energy generation contours. In this figure, convective regions are shaded in
grey, $\epsilon_\nu$ is the specific energy loss rate due to neutrino
production by nuclear reactions and $t^*$ is the time until core collapse. The
turquoise dot-dashed line marks the boundary of the C-free core, which is
defined as the mass coordinate below which the mass fraction of C is lower than
$10^{-4}$. The initial setup of our \textsc{PPMstar} simulations is based on
the structure of the convective O shell in this MESA model at
$\log_{10}(t^*\,/\,\mathrm{yr}) = -1.50$, i.e.  about 10 days before core
collapse (marked with a vertical dashed line). The solid portion of the
vertical line at $\log_{10}(t^*\,/\,\mathrm{yr}) = -1.50$ shows the region of
the star that was simulated in \textsc{PPMstar}.}

\label{fig:O-shell-kip}
\end{figure}

Nucleosynthesis in massive stars is also sensitive to the convective structure
of the core during the advanced burning stages. \citet{Rauscher2002},
reporting on their computations of the complete nucleosynthesis in massive
stars, highlighted an event in a 20~\msun~model in which the convective
O-burning shell and the convective C-burning shell merged into one deep
convective layer, engulfing in addition the Ne-burning shell. In their
computations, this event was responsible for a late-time neutron burst that
resulted in large overproduction of neutron-capture nuclei.  The impact or
likelihood of such a shell interaction has not yet, to our knowledge, been
studied or reported on in any great detail. However,
\citet{Bazan1994,Bazan1998} and \citet{Asida2000} have shown that the
entrainment of \iso{12}{C} from the overlying stable layer into the convective
O-burning shell produced hot spots of nuclear burning which do feed back and
affect the flow. \citet{Meakin2006} and \citet{Arnett2011}, performing 2D
simulations of multiple stratified convective burning shells, reported on the
interaction of the shells by wave propagation in the intershell cavity.

\begin{figure}

\includegraphics[width=\linewidth, clip=true, trim=0mm 1mm 0mm 
                 2mm]{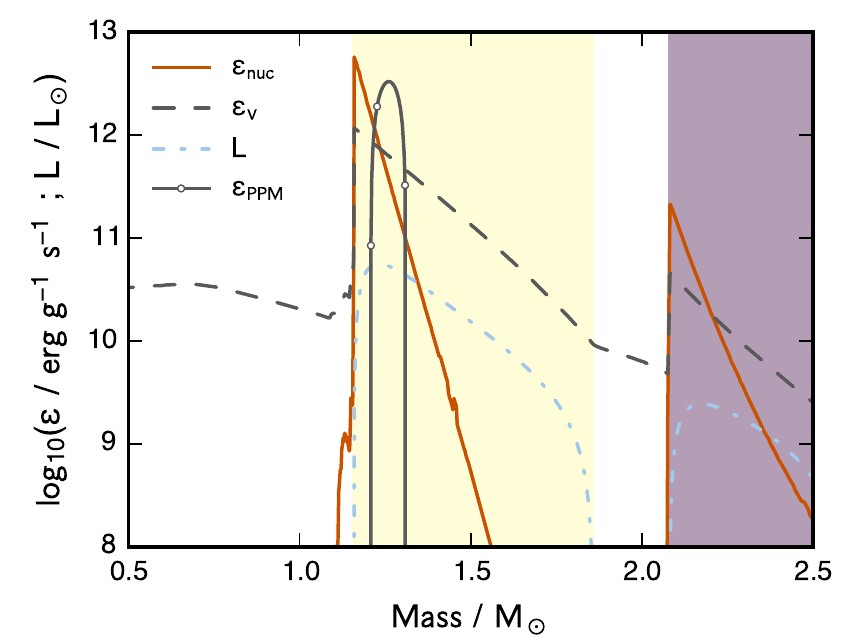}

\caption{Nuclear energy generation rate ($\epsilon_{\rm nuc}$), thermal
neutrino energy loss rate ($\epsilon_\nu$) and luminosity profiles in the
25~\msun~MESA model during shell oxygen burning about 10 days before
core collapse ($\log_{10} t^* = -1.50$ in Figs.~\ref{fig:O-shell-kip} and
\ref{fig:mesa-props}). The oxygen shell reaches from about
1.15~\msun~($4.1\Mm$) to about 1.85~\msun~($8.0\Mm$). The second peak at about
2.08~\msun~($9.4\Mm$) is the base of the convective C shell. The extent of the
$4\pi$ {\sc PPMstar} hydrodynamic simulations, in which the combined effects of
nuclear burning and neutrino losses is approximated by the solid dark grey
curve $\epsilon_{\rm PPM}$, can be seen in Fig.~\ref{fig:P_setup}. The
convectively unstable regions in the underlying MESA model are shaded
(\emph{yellow}: O-burning shell; \emph{purple}: C-burning shell).}

\label{fig:setup_energies}
\end{figure}

The temporally and spatially stochastic fluctuations in the energy generation
rate due to the burning of entrained fuel as it is advected toward the flame at
the base of the convection zone in the simulations of \citet{Arnett2011}
results in an enhanced entrainment rate of the fuel. The enhanced entrainment
rate boosts the nuclear energy generation rate and towards the end of the
calculation, large sloshing motions appear.  A similar phenomenon was reported
by \citet{Herwig2014} in a \textsc{PPMstar} simulation of H ingestion into the
He-burning shell flash during a very late thermal pulse (convective He
shell flash; VLTP) of a post-AGB star.

In this work, idealised simulations of convection in the first O-burning shell
of a massive star are performed with the \textsc{PPMstar} code in $4\pi$
geometry. The original motivation of this work was to investigate the numerical
convergence properties of mass entrainment at the top convection boundary. A
similar investigation for He-shell flash convection was performed by
\citet{Woodward2015}, who found a very low entrainment rate at a seemingly very
stiff convective boundary. In that work it was shown that the numerical
simulations approach convergence for this property. One goal of this paper is
therefore to establish convergence properties for the entrainment rate for this
alternative setup. Following encouraging initial results, the analysis was
extended to develop a mixing model in the diffusion framework that describes
convective mixing in the region of the upper boundary of O-shell convection.

The concept and setup of the idealised simulations are described in
Section~\ref{sec:sims} along with the stellar evolution calculations on which
they are based. The results of the simulations are described in
Section~\ref{sec:results}, including a 1D diffusion analysis connecting the 3D
hydrodynamic simulations back to 1D stellar models.
Section~\ref{sec:discussion} contains a comparison to other works. In
Section~\ref{sec:conclusions} the results are summarized and a brief outlook is
given.

\begin{figure}
\includegraphics[width=\linewidth, clip=true, trim= 0mm 10mm 0mm
                 6mm]{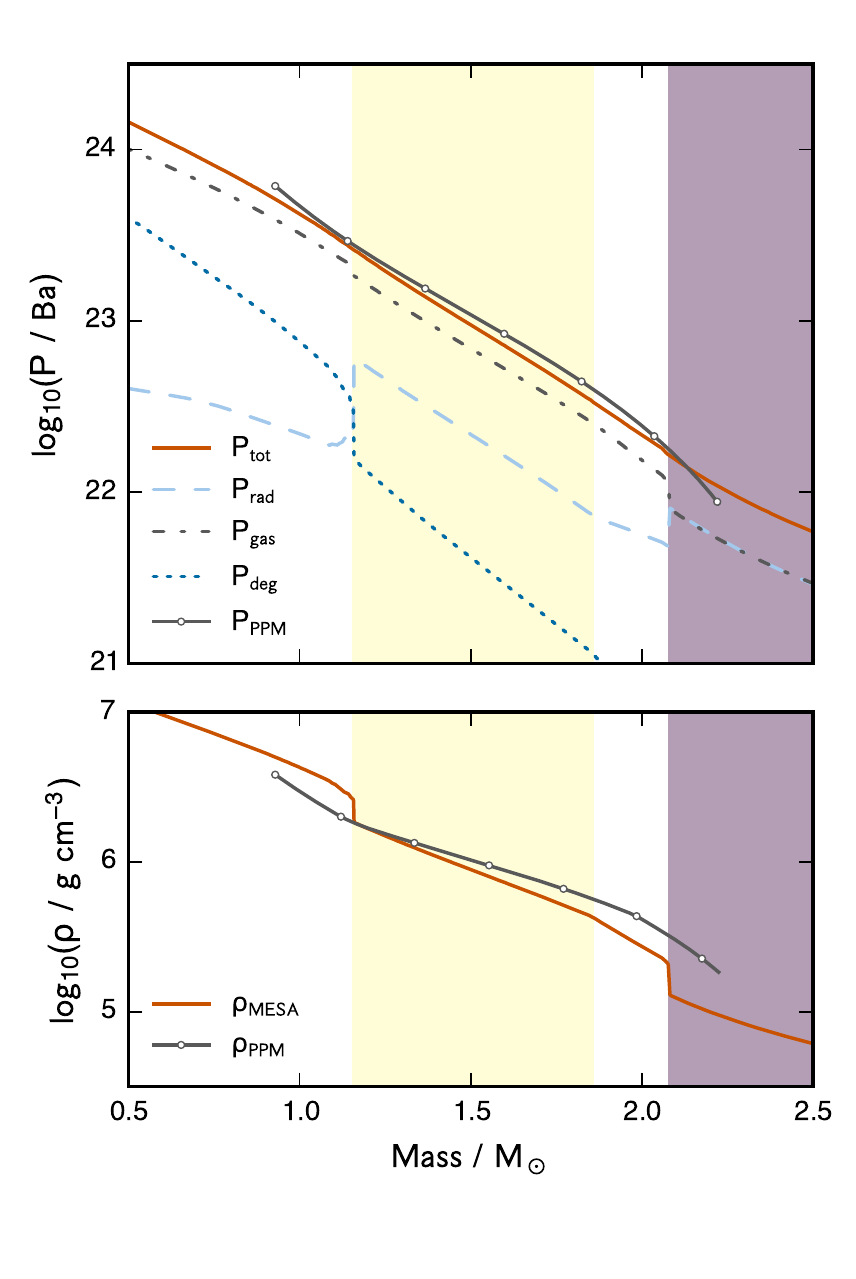}

\caption{Contributions to the total pressure (top panel) and density
stratification (bottom panel) during the shell O-burning phase of the
representative MESA model about 10 days before core collapse ($\log_{10}
t^* = -1.50$ in Figs.~\ref{fig:O-shell-kip} and \ref{fig:mesa-props}). The
initial pressure and density stratifications of the {\sc PPMstar} hydrodynamic
simulations are shown with subscript PPM (solid dotted grey lines). The
convectively unstable regions in the underlying MESA model are shaded
(\emph{yellow}: O-burning shell; \emph{purple}: C-burning shell).}

\label{fig:P_setup}
\end{figure}

\section{Simulations}
\label{sec:sims}

Two types of simulations of the first convective O-burning shell were performed
in the present work: 1D stellar evolution calculations using the MESA code
\citep{Paxton:2010jf,Paxton:2013km,MESA2015} and explicit 3D hydrodynamic
simulations using the \textsc{PPMstar} code \citep{Woodward2015}. As is
described in the following sections, the MESA models boast an impressively
detailed picture of the microphysics while the \textsc{PPMstar} simulations are
performed with idealised physics assumptions. These assumptions are but one of
the reasons that it is possible to use very high numerical resolution in $4\pi$
geometry with the \textsc{PPMstar} code and still follow the simulation for
27~minutes with the high-resolution grid and up to 55~minutes using our
low-resolution grid.  Additionally, we would like to be able to compare the O
shell simulations directly with the AGB thermal pulse entrainment simulations
by \citet{Woodward2015}, and this requires being able to separate the effects
of macrophysics and microphysics. The latter will be investigated in future
work.  This compromise on the details of the microphysics means that the
initial setup of the \textsc{PPMstar} simulations will not exactly match the
spherically symmetric MESA models that they are based on. It is of course
reasonable to expect that the 3D hydrodynamic and 1D hydrostatic models will
differ in many aspects, and in the present work it is assumed that the
important aspects to optimize for are the aspect ratio and driving luminosity
of the convection zone, and the mean molecular weights of the fluids interior
and exterior to the convection zone.  Furthermore, when comparing 1D and 3D
simulations, it can actually become quite challenging to determine metrics by
which the two can actually be compared (see Section~\ref{sec:1d-3d}). We define
the aspect ratio of a convection zone to be $(\rub-\rlb)/\rub$, where
\rub\ and \rlb\ are the radii of the upper and lower boundary of the convection
zone, respectively. While perhaps it may seem more intuitive to have the depth
non-dimensionalised by putting \rlb\ in the denominator, such a definition would
be problematic for convective cores for which $\rlb=0$.

\begin{figure}
\centering
\includegraphics[width=\linewidth, clip=true, trim=0mm 2mm 0mm 
                 5mm]{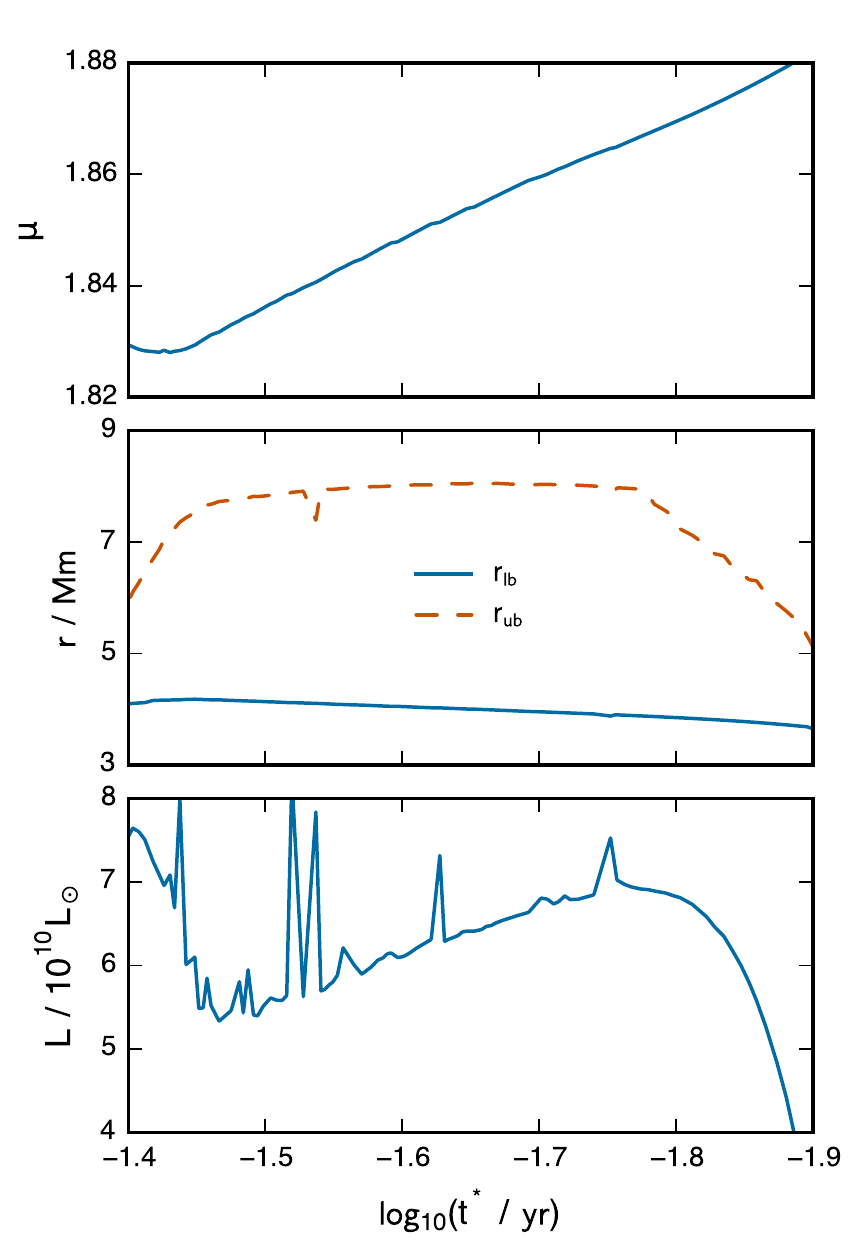}

\caption{Time evolution of the defining quantities of the convective O-burning
shell in a $25~M_\odot$ stellar model computed with the MESA stellar evolution
code. The x-axis is the log of the time remaining until core collapse $t^*$
(cf.~Fig.~\ref{fig:O-shell-kip}); \rlb\ and \rub\ are the radius of the bottom
and top of the convective O-burning shell, $\mu_\mathrm{conv}$ is the mean
molecular weight in the convective region and $L$ is the peak luminosity
driving the convection.}

\label{fig:mesa-props}
\end{figure}

\subsection{MESA simulations}
\label{sec:MESAsimulations}

Stellar evolution calculations of a $25\msun$ non-rotating massive star with
metallicity $Z=0.02$ were performed using the MESA code\footnote{The
initial chemical composition of the stellar models was $X_\mathrm{ini} =
0.706$, $Y_\mathrm{ini} = 0.274$ and $Z=0.02$ with the metal distribution after
\citet{grevesse:93}}. This stellar model is a continuation of the \textsc{m25}
model from \citet{Jones:2015cc} beyond core helium-burning.  The MESA model
assumes hydrostatic equilibrium and uses the mixing length theory of convection
(MLT) with mixing length $\ell=1.6H_\mathrm{P}$, where $H_\mathrm{P}$ is the
pressure scale height.  Convective stability is determined on the basis of the
Schwarzschild criterion and convective mixing of nuclear species is modelled as
a diffusive process.  Inside the convection zone, the diffusion coefficient is
assumed to be $D=\frac{1}{3}v\ell$, where $v$ is the convective velocity
predicted by MLT.  Mixing at the convective boundaries of the oxygen-burning
shell was approximated with an exponentially-decaying diffusion coefficient as
proposed by \citet[][see also \citealp{herwig:97}]{freytag:96}.  The convective
boundary mixing parameter used in this approximation was $f=2\times10^{-3}$,
which is used to give the e-folding length of the diffusion coefficient,
$\frac{1}{2}f_\mathrm{CBM}H_P$.  The nuclear species included in the nuclear
reaction network in the MESA simulations performed as part of this work are
given in Fig.~\ref{fig:MESAnetwork}.  

A Kippenhahn (convective structure evolution as a function of time) diagram of
the $25\msun$ MESA model with positive nuclear energy generation
contours is shown in Fig.~\ref{fig:O-shell-kip}. The figure shows only the
central region of the star (the inner-most $3\msun$) during the core and shell
O-burning phases. A deep overlying convective C shell is present throughout,
although only the bottom portion can be seen in the figure. The initial setup
of our \textsc{PPMstar} simulations is based on the structure of the convective
O shell in this MESA model at $\log_{10}(t^*\,/\,\mathrm{yr}) \approx
-1.5$, i.e.  roughly 10 days before core collapse.  The O-burning shell in the
$25~\msun$ stellar model sits atop a convectively stable silicon core of about
$4\Mm$ in radius (about $1.1~\msun$).  The shell is convective and is driven by
the input of heat into the gas, which is predominantly from the fusion of
\iso{16}{O} nuclei with a small contribution made by other nuclear reactions
and gravitational contraction. The majority of this energy generation is
confined to a thin shell located at a radius of $4.1\Mm$ (about $1.15~\msun$)
where both the temperature and the concentration of fuel are high enough for
fusion to occur. The convection zone extends out from the shell at $4.1\Mm$ to
about $8.0\Mm$ ($1.85~\msun$). Above the convective oxygen-burning shell is a
deep convective C-burning shell.  The two convective shells are separated by a
stable layer of roughly $1.5\Mm$ (about $0.2~\msun$) of material, mostly
composed of O, Ne, Mg and a small amount of Si (i.e. products of incomplete Ne
burning).  The aspect ratio of the convective O-burning shell is thus about
0.5.  In a survey performed as part of this work, massive star models were
computed with the MESA code with initial masses ranging from 12 to $25~\msun$.
The aspect ratio of this first O-burning shell is quite similar in all of the
models across the range of initial masses that were simulated.

The dominant source of pressure in the convective O-burning shell comes from
the gas, which contributes about 75 per cent of the total pressure, while
degenerate electrons contribute at the per cent level. Radiation pressure
accounts for the remaining fraction of the total pressure
(Fig.~\ref{fig:P_setup}). The contributions in the overlying stable layer are
similar to the O-shell, but in the convective C-burning shell (with its base at
a mass coordinate of about $2.1\msun$), the contribution to the pressure is
split equally between the gas and radiation.

Several mechanisms lead to the rapid production of neutrinos in the deep
interior of massive stars during their advanced burning stages \citep[see,
e.g.,][]{Itoh1996}. The densities are, however, low enough that the interaction
of the neutrinos with the stellar plasma is negligible and they stream freely
from the stellar core. The production of neutrinos accelerates the evolution of
massive stars during their post-He core burning phases: nuclear binding energy
must be released at a rate that not only provides the necessary luminosity
to support the star but also at a rate that compensates for the neutrino energy
losses.  Radial profiles of the nuclear energy generation rate
$\epsilon_\mathrm{nuc}$, thermal neutrino loss rate (i.e.~neutrinos produced by
the plasma via pair production, which is by far the dominant source of
neutrinos in the O-burning shell) $\epsilon_\nu$ and luminosity in the vicinity
of the oxygen shell are shown in Fig.~\ref{fig:setup_energies}.  Despite the
rapid neutrino energy sink, the net power generated due to the nuclear burning
is still some $5.2\times10^{10}~\lsun$ ($2\times10^{44}$~erg~s$^{-1}$).

\begin{figure}
\includegraphics[width=\linewidth, clip=true, trim=0mm 0mm 0mm
                 1mm]{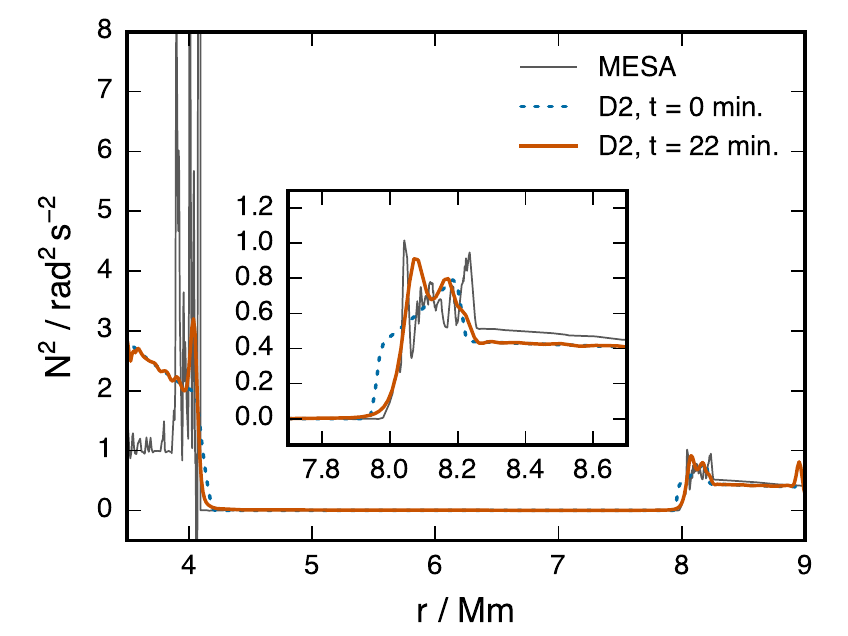}

\caption{Radial profiles of the squared Brunt-V\"ais\"al\"a frequency $N^2$ at
the beginning of the simulation and at the 22$^\mathrm{nd}$ minute in the
\textsc{d2} run. The MESA model is shown for comparison and to illustrate that
the amplitude and slope of the $N^2$ profile at the upper boundary in the
\textsc{PPMstar} model are very similar.}

\label{fig:N2_MESA_vs_PPM}
\end{figure}

\subsection{{\sc PPMstar} simulations}
\label{sec:PPM-setup}

Hydrodynamic simulations of O-shell convection were performed in 3D with $4\pi$
geometry using the {\sc PPMstar} code. The code as well as its performance for
a similar application (He-shell flash convection in a low-mass star) are
described in detail in \citet{Woodward2015}. The setup procedure, as well as
the physics assumptions in this work are the same as in the He-shell flash
convection simulations. In summary, an ideal gas equation of state is adopted
and a geometrically representative stratification is produced by combining
three piece-wise polytropes. Particular attention is paid to reproducing the
pressure and density stratification. These are the important quantities for
describing the flow; the temperature, if it were considered, would be
over-estimated since the contribution of the radiation to the total pressure
(see Fig.~\ref{fig:P_setup}) is assumed to be provided solely by the gas.  In
Fig.~\ref{fig:mesa-props} defining quantities of the convective O-burning shell
in the 25~\msun~MESA model -- the mean molecular weight in the convection zone
$\mu$, the radial coordinate of the convective boundaries \rub and \rlb, and
the peak luminosity inside the convection zone $L$ -- are given as a function
of time until the onset of core collapse (cf.~Fig.~\ref{fig:O-shell-kip}). This
figure is intended to demonstrate the similarity of the initial setup of the
PPMstar simulations presented in this work to the structure one would expect
the O-burning shell to have when much more detailed microphysics have been
considered.

\begin{figure}
\includegraphics[width=\linewidth, clip=true, trim=0mm 0mm 0mm
                 1mm]{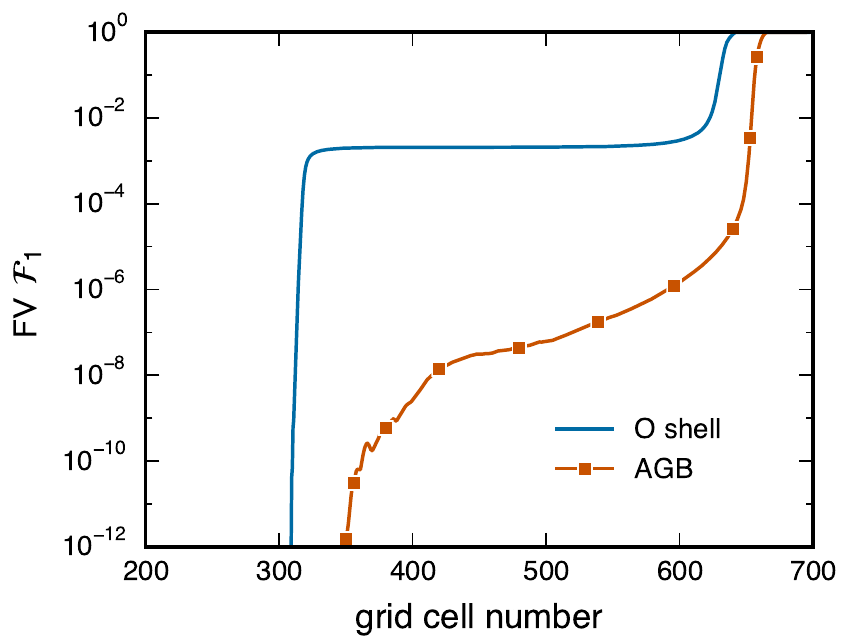}

\caption{Fractional volume profile for He-shell \citep{Woodward2015} at time
step 808672 (AGB) and O shell ({\sc d2} from this work) runs at the same
product of number of time steps and Courant number (i.e. the same computational
effort). The much larger amount of material that has been entrained in the O
shell simulation compared with the AGB simulation after expending the same
amount of computational effort is the reason why O-burning shell is well suited
for studying the convergence properties of {\sc PPMstar} simulations of stellar
convection.}

\label{fig:FV_comparison_at_same_tstep}
\end{figure}

\begin{table*} 
\centering
\begin{tabular}{llrrrrrrr}
\toprule
id &
grid~$^a$ &
$t_\mathrm{sim}~^b$ &
\centering $L~^c$ &
\centering $\dot{M}_{\rm e}~^d$ &
\centering $\dot{r}_{\rm ub}~^e$ &
\centering $v_\mathrm{r}~^f$ &
\centering $v_\perp~^g$ &
\centering $M~^h$

\tabularnewline
 &
 &
\centering (min) &
\centering ($10^{10}L_\odot$) &
\centering ($10^{-6}~M_\odot/{\rm s}$) &
\centering (km/s) &
\centering (km/s) &
\centering (km/s) &
$10^{-2}$
\tabularnewline
\midrule
\textsc{d1}  & $768^3$  &  55.2 & 11.8 & 1.15 & 0.016 & 32.0 & 38.0 & 1.21 \\
\textsc{d2}  & $1536^3$ &  27.3 & 11.8 & 1.33 & 0.018 & 32.6 & 36.8 & 1.19 \\
\textsc{d8}  & $768^3$  &  36.2 & 29.5 & 3.60 & 0.035 & 44.4 & 50.6 & 1.62 \\
\textsc{d5}  & $768^3$  &  37.2 & 59.1 & 8.07 & 0.060 & 53.5 & 64.1 & 2.08 \\
\textsc{d6}  & $768^3$  &  41.3 & 118  & 16.8 & 0.116 & 69.4 & 87.7 & 2.82 \\
\textsc{d9}  & $768^3$  &  43.5 & 295  & 38.3 & 0.257 & 92.6 & 116  & 3.77 \\
\textsc{d10} & $768^3$  &  43.7 & 591  & 79.4 & 0.524 & 123  & 151  & 4.85 \\
\bottomrule
\end{tabular}

\begin{tabular}{lll}

$^a$~grid resolution & $^b$~simulated time & $^c$~driving luminosity \\
$^d$~entrainment rate & $^e$~upper boundary velocity & $^f$~maximum radial
velocity \\
$^g$~maximum tangential velocity & $^h$~maximum Mach number &

\end{tabular}

\caption{Summary of the hydrodynamic simulations that have been performed with
the {\sc PPMstar} code for this work, ordered by driving luminosity. {\sc d1}
and {\sc d2} are both representative of the actual conditions in the first
O-burning shell in a $25~\msun$ star (c.f. Fig.~\ref{fig:mesa-props}) and the
other simulations constitute a heating series in which the properties of
convection and entrainment are studied as a function of the luminosity driving
the convection. The entrainment rate $\dot{M}_\mathrm{e}$ is given for the
upper convective boundary. $v_{\rm r}$ is the maximum tangential component of
the velocity of the fluid in the upper half of the convection zone.}

\label{tab:run-info}
\end{table*}

\begin{figure*}
\includegraphics[width=\linewidth]{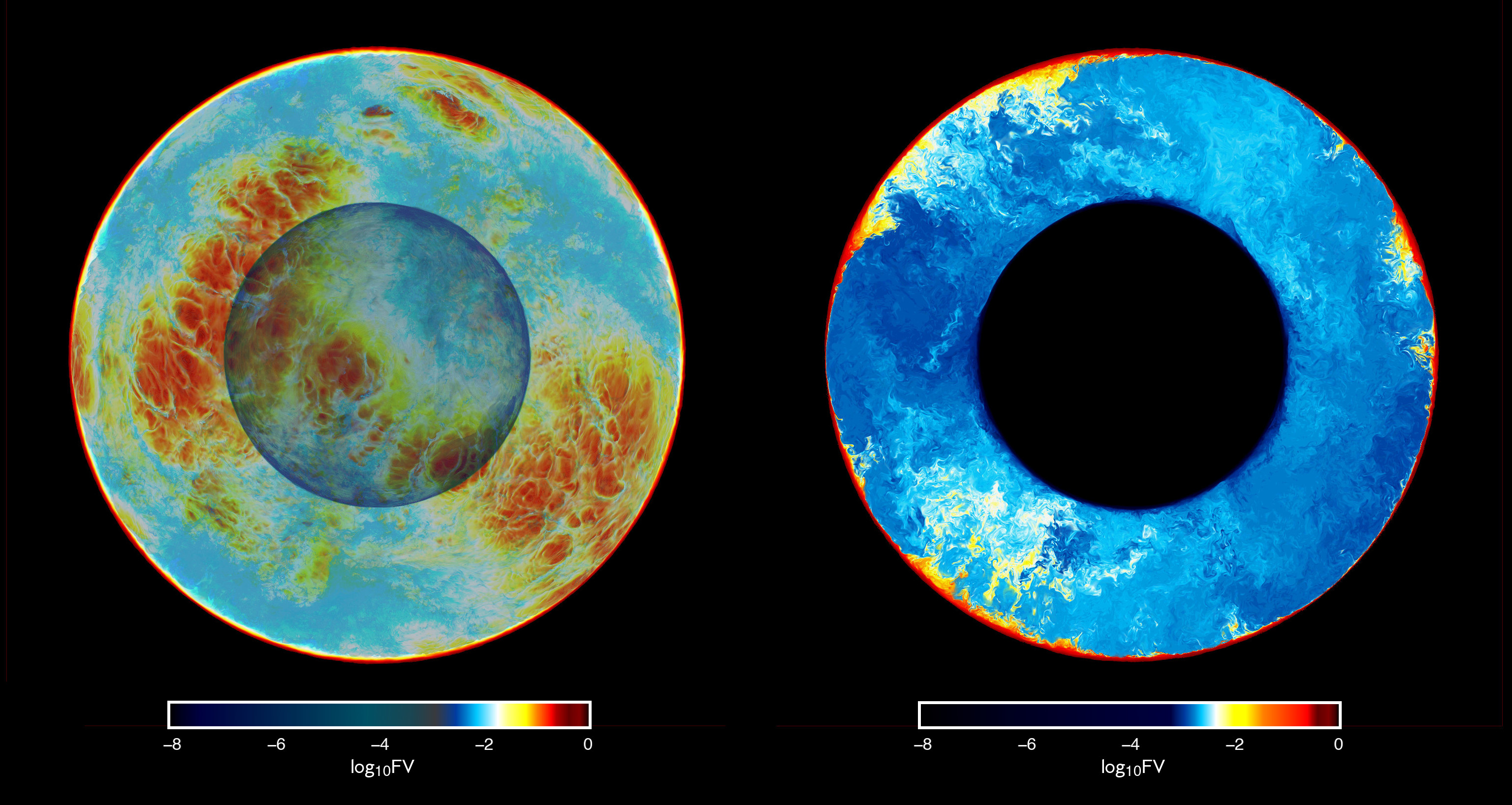}

\caption{Fractional volumes of the entrained fluid in the 1536$^3$
\textsc{PPMstar} run {\sc d2} at 25.7 minutes of simulated time (dump number
154). The left image is a projection of the far hemisphere of the simulation;
the near hemisphere has been cut away and the core has been made almost
transparent, but is still visible as a faint purple sphere. The right
image is a thin slice through the sphere. In both images, fluid that began
inside the convection zone has been made transparent and therefore only
entrained fluid is visible. That entrained fluid has a fractional volume that
is indicated by the colour scale. See Section~\ref{sec:PPM-props} of the text
for a detailed description.}

\label{fig:1536-FVs}
\end{figure*}

The convective shell is represented by a polytropic stratification with $\gamma
= 5/3$. The layers above and below have stable stratifications with $\gamma_a =
1.35$ and $\gamma_b=1.05$, respectively. The convection is driven by a
constant-volume heating rate of $1.18\times10^{11}~\lsun$, which is equivalent
to the rate of energy generation due to oxygen burning in the MESA model to
within a factor of two (see Fig.~\ref{fig:mesa-props} and
Section~\ref{sec:MESAsimulations}) accounting for the neutrino energy losses
from both nuclear reactions and pair production. The heated shell has a
thickness of $5\times10^7$~cm (Fig.~\ref{fig:setup_energies}) and the
transition region between the top of the convective shell and the overlying
stably stratified layer is $2.5\times10^7$~cm (about $\frac{1}{6}H_P$), centred
at about $8.08\times10^8$~cm. The thickness of the transition layer was chosen
to give the best agreement between the profiles of $P$ and $\rho$ in the
piecewise polytropic {\sc PPMstar} setup and the MESA model on which it was
based. It is important to also note that the thickness of the transition layer
should be such that it can be adequately resolved on both the $768^3$ and
$1536^3$ resolutions grids of {\sc PPMstar}. At the base of the convective
shell, the density and gravitational acceleration are
$1.82\times10^{6}$~g~cm$^{-3}$ and $9.02\Mm~\mathrm{s}^{-2}$, respectively. At
the top convection boundary a pressure scale height is
$H_\mathrm{P}=1.5\mathrm{Mm}$. 

At the start of the simulation, the stable layer consists of
fluid $\mathcal{F}_1$ with a mean molecular weight $\mu_1 = 1.802$, and the
rest of the simulation domain is filled with fluid $\mathcal{F}_2$ with a mean
molecular weight $\mu_2 = 1.848$. The resulting profile of the squared
Brunt-V\"ais\"al\"a frequency\footnote{This expression is equivalent to the
more familiar expression $N^2 = (g/H_p)(\nabla_\mathrm{ad} - \nabla +
\nabla_\mu)$ for an ideal gas, in which \mbox{$\nabla - \nabla_\mu = 1 -
\nabla_\rho$}.} $N^2 = (g/H_p)(\nabla_\mathrm{ad} - 1 + \nabla_\rho)$ as
compared with that given by MESA is shown in Fig.~\ref{fig:N2_MESA_vs_PPM}. The
MESA model in Fig.~\ref{fig:N2_MESA_vs_PPM} matches the position of the
convective boundary in {\sc PPMstar}, although the mean molecular weight of the
convective fluid in the MESA model is slightly too low at $\mu = 1.842$. The
$N^2(r)$ profile in a somewhat more evolved MESA model (not shown in
Fig.~\ref{fig:N2_MESA_vs_PPM} for clarity) that matches $\mu_2 = 1.848$ of the
{\sc PPMstar} model has a similar structure and amplitude, but the top of the
convection zone in that model is located at a radius $\sim 0.1\,\mathrm{Mm}$
larger than in the {\sc PPMstar} model.

\begin{figure*}
\includegraphics[width=\linewidth]{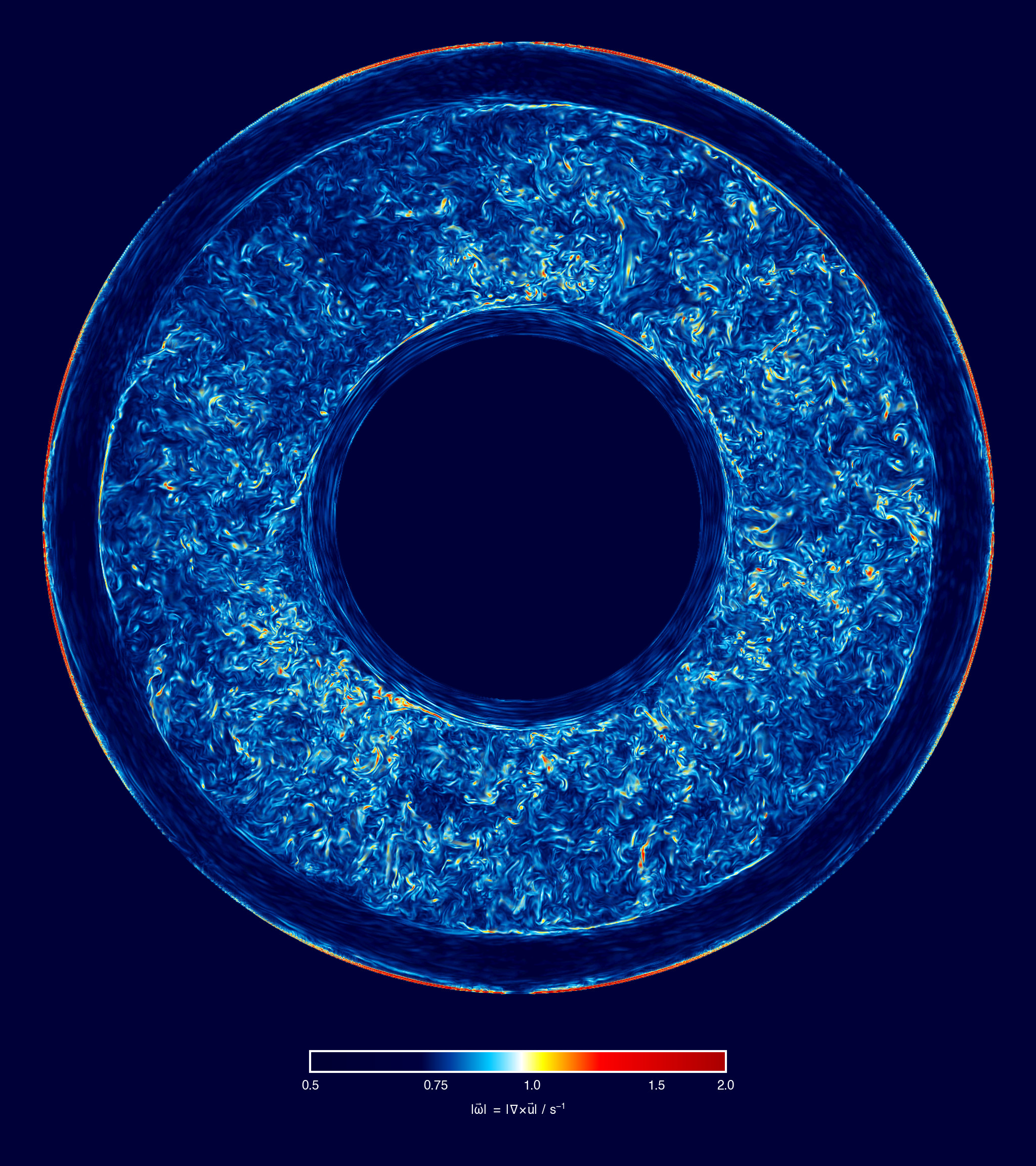}
\caption{A volume rendering of the vorticity of the fluid in the $1536^3$
\textsc{PPMstar} run \textsc{D2} at $25.7\mathrm{min}$ of simulated time (dump
number 154). From the outside inward the domain boundary, a stable layer, the
convectively unstable region, a rather thin stable region below the turbulent
convection zone and the inert core can be distinguished.}
\label{fig:1536-vort}
\end{figure*}

The O-burning shell convection setup is very similar in important respects to
the VLTP convection in Sakurai's object, which has also been simulated using
\textsc{PPMstar} \citep{Herwig:2011dj,Herwig2014}.  The aspect ratios (see the
beginning of Section~\ref{sec:sims} for a definition) of the O-burning
convective shell and the VLTP shell flash convection zone are 0.49 and 0.69,
respectively.  Mach numbers in the convection flows are also similar.  An
appreciable difference is the ratio of the mean molecular weight of the
convective fluid to that of the overlying stable fluid.
The VLTP in Sakurai's object has such a mean molecular weight ratio of 2.26,
while in the O-burning shell convection problem the same ratio is only 1.02.
Another difference is that in these O-shell simulations we do not take into
account the burning of any entrained material from above. In that regard the
simulation approach and goals are the same as those of the entrainment
simulations for He-shell flash convection of \citet{Woodward2015}. In this
paper we wish to study the entrainment and general flow properties without the
additional effect of possible feedback from nuclear burning of the entrained
material. A summary of the 3D simulations performed for this work is given in
Table~\ref{tab:run-info}.

\section{Results}
\label{sec:results}

\subsection{General properties of the flow}
\label{sec:PPM-props}

The 3D simulation begins with a constant entropy in the convection zone that is
driven into an unstable convection flow by introducing a continuous injection
of heat with a spherically symmetric profile near the base of the convection
zone (see Fig.~\ref{fig:setup_energies}).  Rather than stir the flow initially
with a spectrum of disturbances, we instead allow the very slight differences
in the numerical representation of the flow on our fine Cartesian grid to
provide the initial perturbations that permit some of the gas to rise and some
to sink in the convection zone.  This approach is not only convenient, but it
also provides a natural way to test that the grid that we use does not
ultimately affect the disturbance spectrum that develops, except of course to
truncate it at its short wavelength end.  This same approach was used in
\citet{Woodward2015}, where visualisations near the beginning of the simulation
(e.g.~their Fig.~6) illustrate the initial grid imprint effects as well as the
process by which they ultimately are overwhelmed by the flow's natural
development of its turbulent convection spectrum.  One might think that waiting
for the initial perturbations to be overwhelmed in this way could be shortened
by beginning with a full spectrum of disturbances, but we would argue that one
must wait this same length of time in the simulation in any case to be sure
that the spectrum is the one the star prefers rather than just the one that has
been put in initially by fiat.

\begin{figure}
\includegraphics[width=\linewidth]{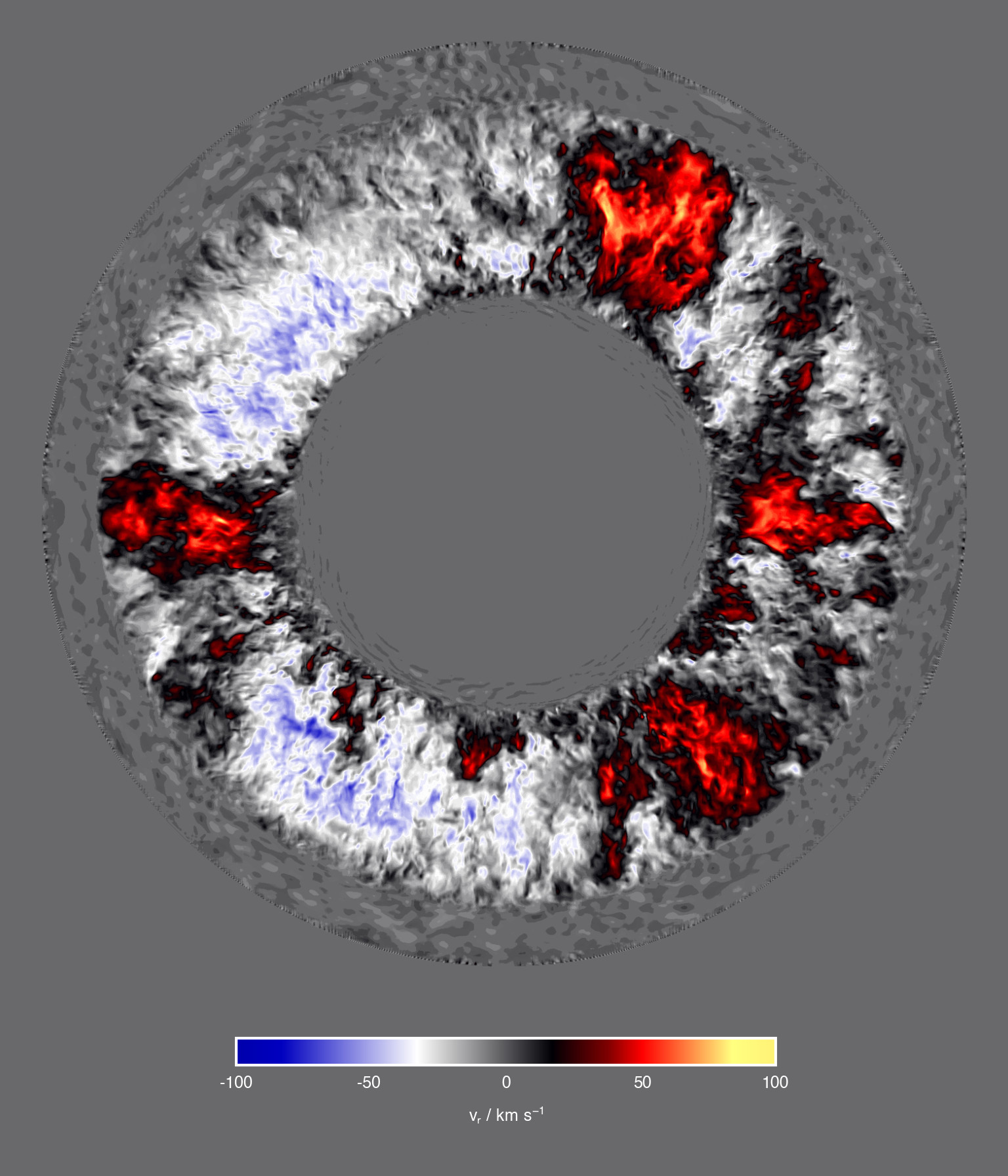}
\caption{Radial velocity of the entrained fluid in the 1536$^3$
\textsc{PPMstar} run {\sc d2}. See Section~\ref{sec:PPM-props} of the text for
a detailed description.}
\label{fig:1536-vertvel}
\end{figure}

In the simulation here ({\sc d2}), the imprint of the grid on the properties of
the flow persists at a detectable level through to about 1300~s (21.7~min), by
which time the flow seems to have been fully turbulent already for a
considerable period. In this setup grid imprints on the disturbance spectrum
persist for a longer time compared to the He-shell convection simulations. The
reason seems to be that the aspect ratio of the convection zone (see
Section~\ref{sec:sims}), about one half, prefers the largest convection cells
that correspond to the Cartesian grid's natural mode. This can be seen at early
times by the prominence of 4 large convection cells in thin sections taken
through the center of the star and aligned with planes of the grid (see movies
at URLs given in Section~\ref{sec:conclusions}).  These develop at the outset
and gradually fade in importance.  Ultimately a still larger mode with just 3
rather than 4 cells visible in such sections develops.  This mode cannot
possibly be preferred by the Cartesian grid, and its development and ultimate
dominance indicates that the flow has taken on a character by 1300~s that
is determined by the physics and the properties of the stellar convection setup
rather than by our numerical treatment.  We choose to use a Cartesian grid
because this makes the design and execution of the simulation code fit the
properties of the machine especially well.  The ultimate overwhelming of grid
imprints indicates that our grid is fine enough that we are justified in
designing the numerics to fit the machine rather than to fit the geometry of
the star.  This analysis of the evolution and eventual disappearance of grid
imprints complements the analysis of simulation properties under grid
refinement (Section~\ref{sec:entrainment}) and provide some level of code
verification for this problem. 

\begin{figure} \includegraphics{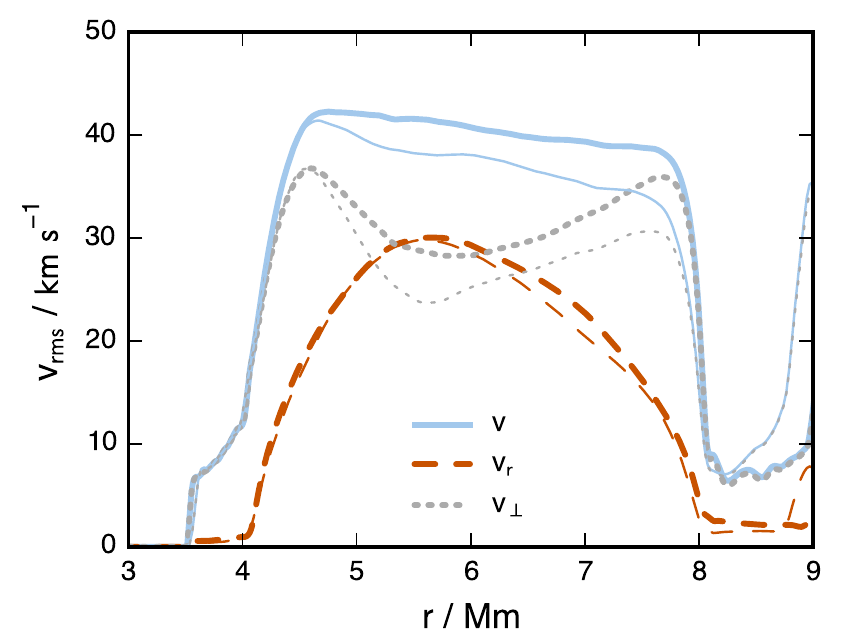}
\caption{Spherically-averaged total, radial and tangential fluid velocity
components after 22 minutes of simulated time (dump number 132) in the {\sc d2}
(thick lines) and {\sc d1} (thin lines) simulations (see
Table~\ref{tab:run-info}).}
\label{fig:vprof-22min}
\end{figure}

The character of the flow above the O-burning shell in this problem is quite
similar to that which we have seen in the shell-flash convection zone during
the very late thermal pulse (VLTP) of post-AGB star Sakurai's object
\citep{Herwig2014} and that of a thermal pulse in an AGB star
\citep{Woodward2015}. The most significant difference from a fluid dynamics
perspective is the relatively large entrainment rate of gas from above the
convection zone that results here for reasons that are explained in
Section~\ref{sec:entrainment}. This is illustrated in
Fig.~\ref{fig:FV_comparison_at_same_tstep}. Fractional volume profiles of the
entrained fluid are shown for the O shell and AGB thermal pulse simulations at
equivalent computational effort, which is defined here as the product of the
number of time steps and the Courant number. The much greater amount of
material that has been entrained into the O-burning shell convection zone
compared to the AGB He shell convection zone after expending the same amount of
computational effort makes the O-burning shell better suited for studying the
properties of convergence in {\sc PPMstar} simulations of stellar convection.
Otherwise, in all these three cases (VLTP, AGB thermal pulse and now O shell)
we have a dominance of very large convection cells, demanding a global
treatment, and low, but not too low, flow Mach numbers, as has been pointed out
earlier. In the left panel of Fig.~\ref{fig:1536-FVs}, we see a hemisphere of
the star, with the hemisphere in the foreground cut away.  The plane facing us
is the $z=0$ plane in the grid of the simulation. The gas of the degenerate
core has been made transparent, and only mixtures of convection zone gas with
entrained gas from above the convection zone are visible. These mixtures at the
bottom of the convection zone have been made to appear dark blue and
semi-transparent, so that the location of this spherical surface is apparent.
High concentrations of entrained gas are red, and as the concentrations become
smaller colors go from red to yellow, white, turquoise, and dark blue,
with differing degrees of opacity.  The pervasive low concentrations of
entrained gas in the convection zone have been made transparent, so that the
convection cell pattern and downward moving entrainment lanes near the top of
the convection zone can be easily seen. In this hemisphere, there are clearly
three large areas where updrafts in large convection cells scrape against the
bottom of the stably stratified layer to reveal the red and yellow colors of
what is a relatively sharp transition between convection zone and stably
stratified gases.  There are also three large regions of descending gas that
include higher concentrations of entrained gas from above the convection zone.
These appear as turquoise and blue.  Where this layer of entraining gas
is cut by our slicing plane at $z=0$, its internal strong gradient of entrained
concentration can be made out.  As discussed in \citet{Woodward2015}, there is
resistance to entrainment where the gas of the convection zone flows roughly
tangentially to the top surface of the convection zone.  Those regions show up
as red and yellow in this image.  In the regions that show up as
turquoise and blue, convection cells meet and the flow is forced to
descend.  It is here that the entrainment happens.  It is resisted by the
relative buoyancy of the entrained gas, but its concentration is not large, and
this resistance is overcome even by the low Mach numbers that apply in this
region.

\begin{figure}
\includegraphics{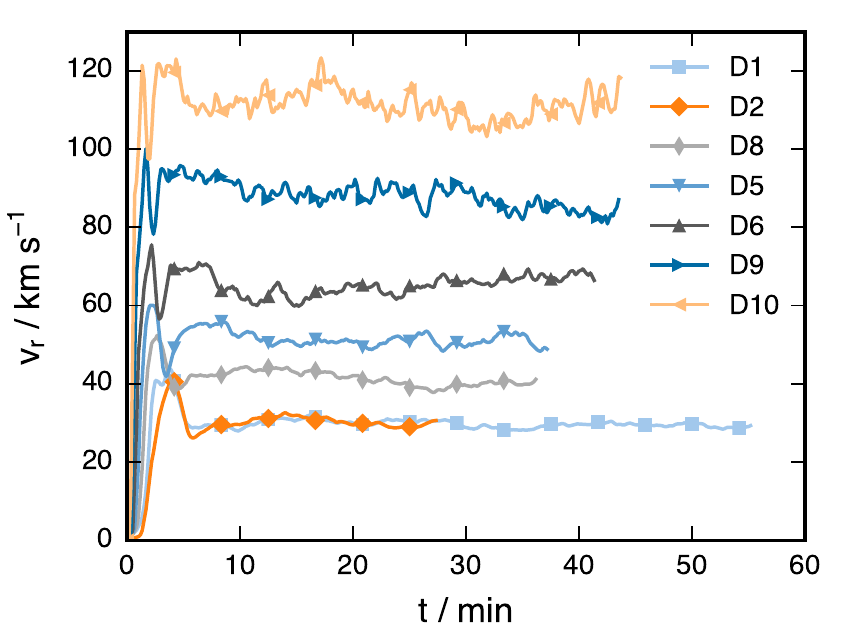}
\caption{Temporal evolution of the maximum RMS radial velocity $v_r$ in the
convection zone for all of the runs listed in Table~\ref{tab:run-info}. All of
the runs experience an initial transient event after which the convection
settles down into a steady state. The {\sc d1} and {\sc d2} simulations are in
very good agreement following their respective initial transients. }
\label{fig:vr_evolution}
\end{figure}

The image on the right in Fig.~\ref{fig:1536-FVs} is the concentration of
entrained gas in the same $z=0$ plane at the same time in the simulation, but
the rendering has been altered to show the entrained concentration not only
near the top of the convection zone but all the way down to its bottom. This is
a thin slice, just 1 per cent of the diameter of the simulation volume in
thickness.  It is clear from this image that globs of relatively enriched gas
descend right to the bottom of the convection zone in the very large, dominant
convection cells that develop due to the convection zone's large depth as a
fraction of the radius of its top surface. In the real star this entrained
material would of course be burned by nuclear reactions, which we ignore in
this set of simulations. The active entrainment near this top surface can also
be seen in the Kelvin-Helmholtz trains of vortices that develop where the flow
near the top of the convection zone turns downward, as was discussed in detail
in \citet{Woodward2015}.  These trains of vortices show up mainly as red and
yellow features in this image. The images in Figs.~\ref{fig:1536-vort} and
\ref{fig:1536-vertvel} are renderings of the vorticity and the radial fluid
velocity, respectively, in this same thin slice through the star. The vorticity
image emphasizes the smallest vortices and shear layers, making clear that a
full, turbulent spectrum of vortices has developed by this time. The radial
velocity image shows the three major upwellings in shades of red and yellow,
while descending gas appears in shades of grey tending to blue in the most
rapidly descending plumes.

\begin{figure}
 \includegraphics[width=\linewidth, clip=true, trim=0mm 3mm 0mm
                  2mm]{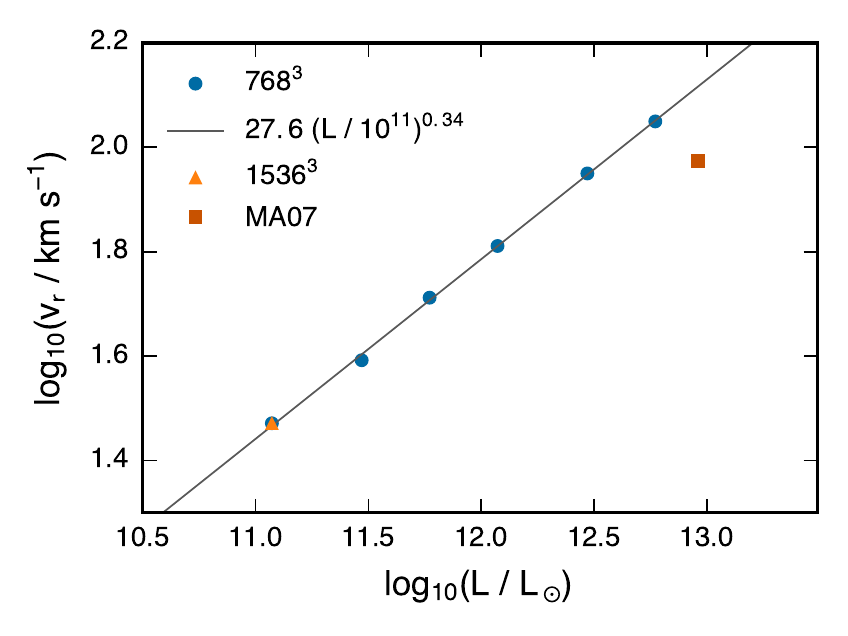}

\caption{Dependence of the maximum RMS radial velocity $v_\mathrm{r}$ on the
luminosity $L$. The RMS velocity profiles were averaged over $10$ data dumps
(spanning $100$\,s of simulation time) centred in the middle of the time
interval, in which the mass entrainment rate was measured for the given run.
The fitting relation, which assumes the same units as the axes of the plot, was
constructed using the $768^3$ runs only. The $1536^3$ run \textsc{d2} and the
3D simulation of \citet{Meakin2007a} are shown for comparison.}

\label{fig:radial_velocity_vs_luminosity}
\end{figure}

\subsection{Velocity profiles}
\label{sec:velocity-profiles}

The spherically-averaged RMS velocity profiles in the \textsc{d2} run are shown
in Fig.~\ref{fig:vprof-22min}. The typical RMS total velocity is $\sim
40$\,km\,s$^{-1}$ with about equal contributions from the radial and tangential
components at $r = 5.6$\,Mm where the radial component reaches a maximum. The
broad local maxima of the tangential velocity $v_\perp$  correspond to the fact
that the speed becomes mostly tangential near the convection zone boundary as
the flow turns around. While both the tangential and the radial velocity
profiles have prominent features the total velocity profile is rather flat.
Velocity profiles in the rest of the runs listed in Table~\ref{tab:run-info}
have the same structure. However, the detailed shape of profiles depends on the
shape of the heat source, see the discussion in
Sect.~\ref{sec:discussion}. Fig.~\ref{fig:vr_evolution} shows that
all of the runs reach a steady state after a short initial transient.

\begin{figure*}
 \includegraphics[width=0.49\linewidth, clip=true, trim=0mm 2mm 0mm
                  2mm]{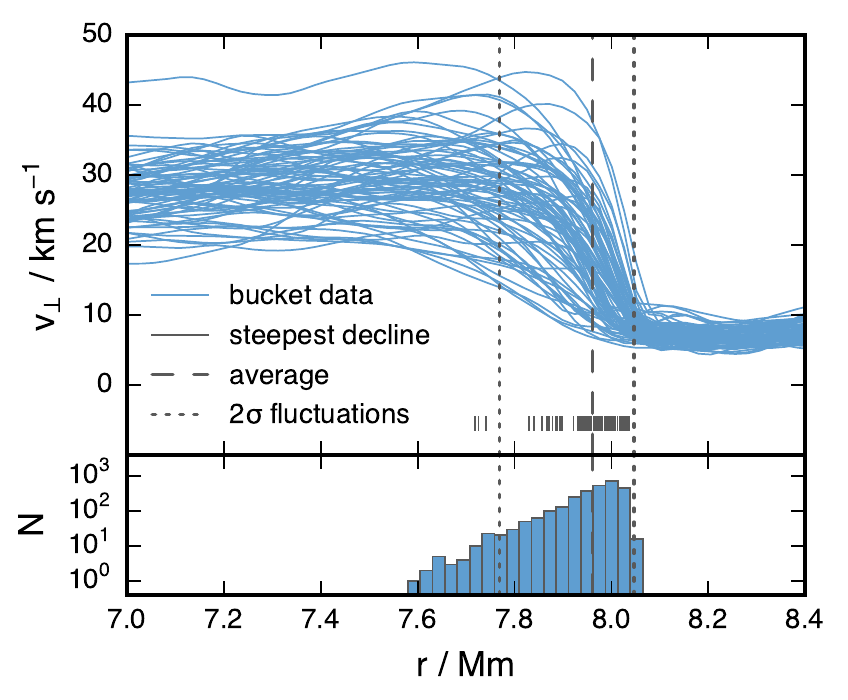}
 \includegraphics[width=0.49\linewidth, clip=true, trim=0mm 2mm 0mm
                  2mm]{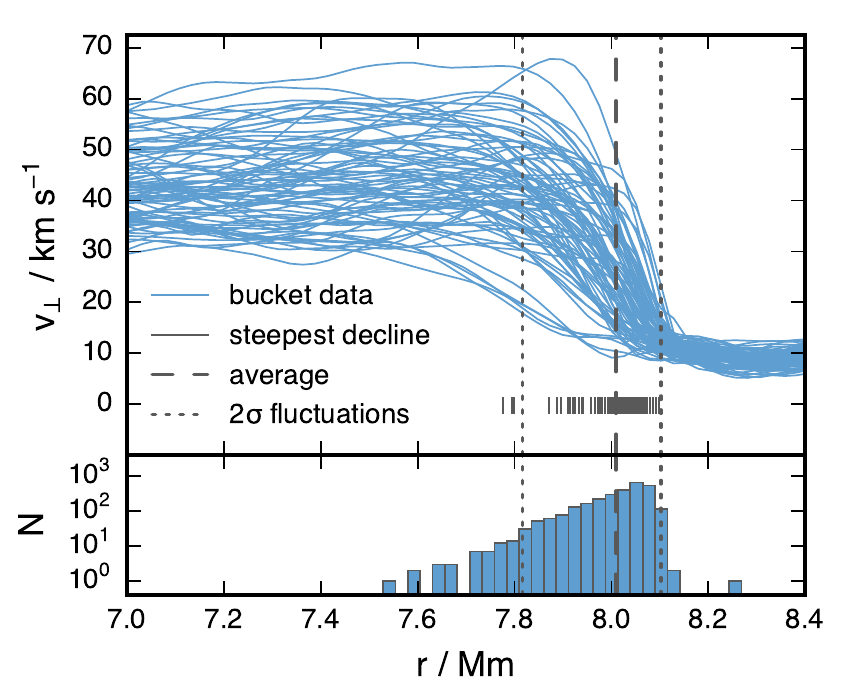}

\caption{RMS tangential velocity profiles at the upper convective boundary in
the \textsc{d1} (left panel) and \textsc{d8} (right panel) runs extracted at $t
= 18\,\mathrm{min}$ from 80 radial \emph{buckets} (see
Fig.~\ref{fig:bucket_map_r_ub} and Section~\ref{sec:1d-3d}). The radius
$\rub$ of the steepest decline in $v_\perp$, which is used to define
the position of the convective boundary, is marked for each profile separately
by a short vertical line. The long, dashed and dotted vertical lines show
the average value of and $2\sigma$ fluctuations in $\rub$, respectively.
The standard deviation $\sigma$ was computed separately for positive and negative
fluctuations. The lower part of each plot shows the distribution of
$\rub$ values collected from 35 data dumps in the time interval from
$t = 15\,\mathrm{min}$ to $t = 20\,\mathrm{min}$. The width of the histogram
bins corresponds to the cell size of the underlying \textsc{PPMstar}
computational grid.}

\label{fig:upper_boundary_vt}
\end{figure*}

\begin{figure*}
 \includegraphics[width=0.49\linewidth, clip=true, trim=0mm 2mm 0mm
                  2mm]{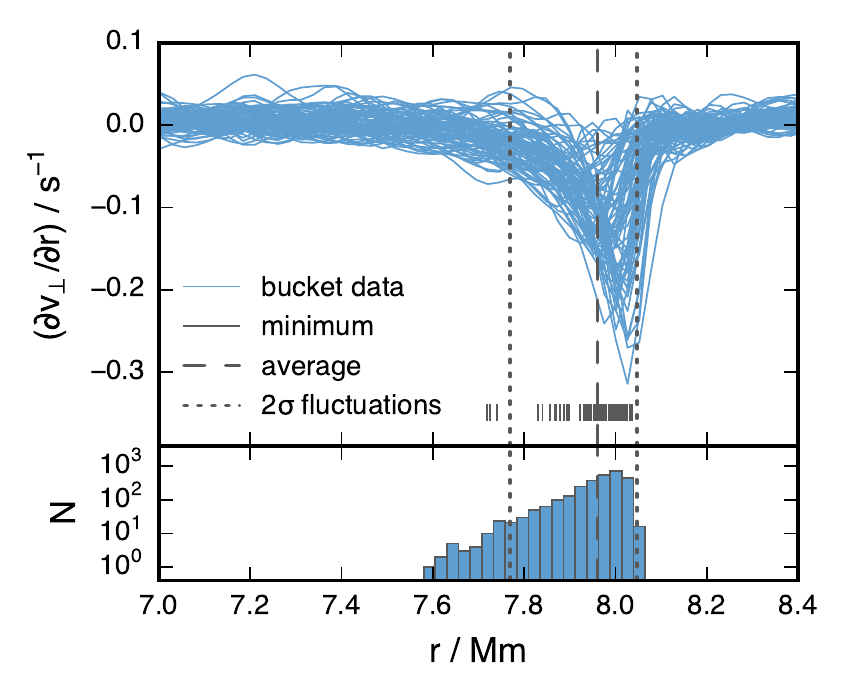}
 \includegraphics[width=0.49\linewidth, clip=true, trim=0mm 2mm 0mm
                  2mm]{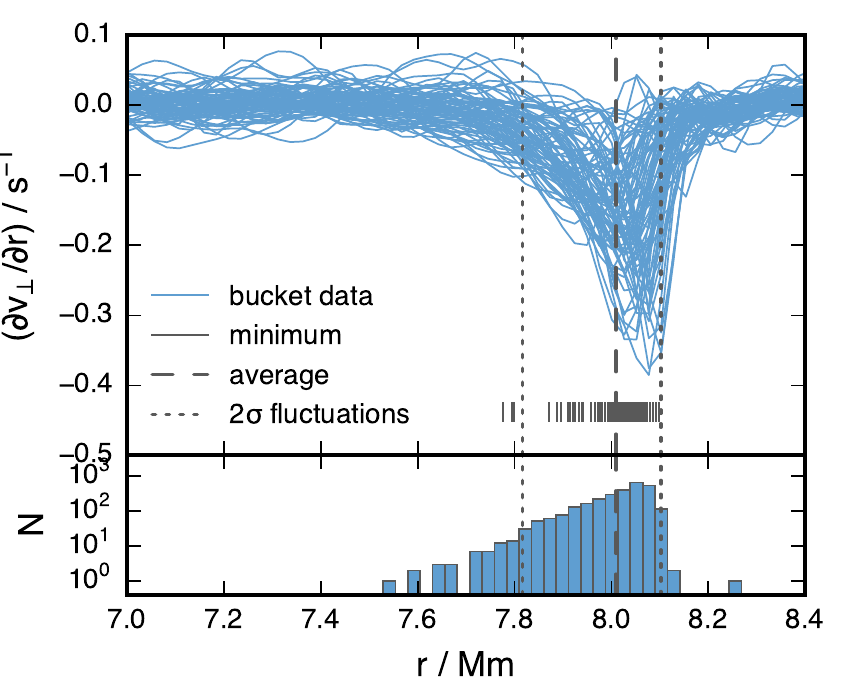}

\caption{As Fig.~\ref{fig:upper_boundary_vt}, but $\partial v_\perp / \partial
r$ is shown for the \textsc{d1} run in the left panel and for the \textsc{d8}
run in the right panel.}

\label{fig:upper_boundary_dvtdr}
\end{figure*}

The flow velocity is closely related to the luminosity $L$ of the heat source.
The best-fit scaling relation shown in
Fig.~\ref{fig:radial_velocity_vs_luminosity} is very close to
\mbox{$v_\mathrm{r} \propto L^{1/3}$}, in agreement with \citet{Porter:2000eo}.
This scaling can be motivated as follows.

In a stationary state, the luminosity $L(r)$ can be decomposed to the fluxes
$F_h$ and $F_k$ of enthalpy and kinetic energy, respectively,
\begin{equation}
L(r) = 4\pi r^2 (F_h + F_k),
\label{eq:total_energy_balance}
\end{equation}
because we neglect the small radiative contribution. Both terms on the right
hand side of Eq.~\ref{eq:total_energy_balance} can be shown to scale in
proportion to $v^3$. The kinetic energy flux scales with the kinetic energy
density and velocity, hence $\mathcal{F}_\mathrm{k} \propto \rho v^3$. The
enthalpy flux is $F_h = \rho v c_p T^\prime$, where $T^\prime$ is a typical
temperature fluctuation between the upflows and downflows. Assuming that the
relative temperature fluctuations $T^\prime/T$  are of the same order as the
relative dynamic pressure fluctuations $p^\prime/p \approx \rho v^2 / p$, we
have $T^\prime \approx T \rho v^2 / p \propto v^2$, because $p \propto \rho T$
by the ideal gas law. The enthalpy flux is then $F_h \propto \rho v^3$.

Another way to see why $L \propto v^3$ is to consider the kinetic energy
generation and dissipation rates in a convection zone. Per unit of volume, the
turbulent dissipation rate is $\epsilon_\mathrm{d} \propto \rho v^3 / l$, where
$l$ is an integral length scale of the turbulence. Therefore also the total
dissipation rate in the convection zone, which in a stationary state is of the
same order as $L$ \citep[see][]{Viallet:2013gz}, scales in proportion to $v^3$.
A similar argument was made by \citet{Biermann1932}.

\subsection{Where is the convective boundary?}
\label{sec:1d-3d}

Standard stellar evolution models, such as our MESA models, are
spherically symmetric. The local Schwarzschild or Ledoux criterion is used to
determine convectively unstable regions. The mixing-length theory (MLT) is then
applied to these regions to estimate the temperature gradient, convective
velocity and convective energy flux. According to MLT, the convective velocity
vanishes at convective boundaries.

\begin{figure*}
 \includegraphics[width=0.49\linewidth, clip=true, trim=0mm 1mm 0mm
                  2mm]{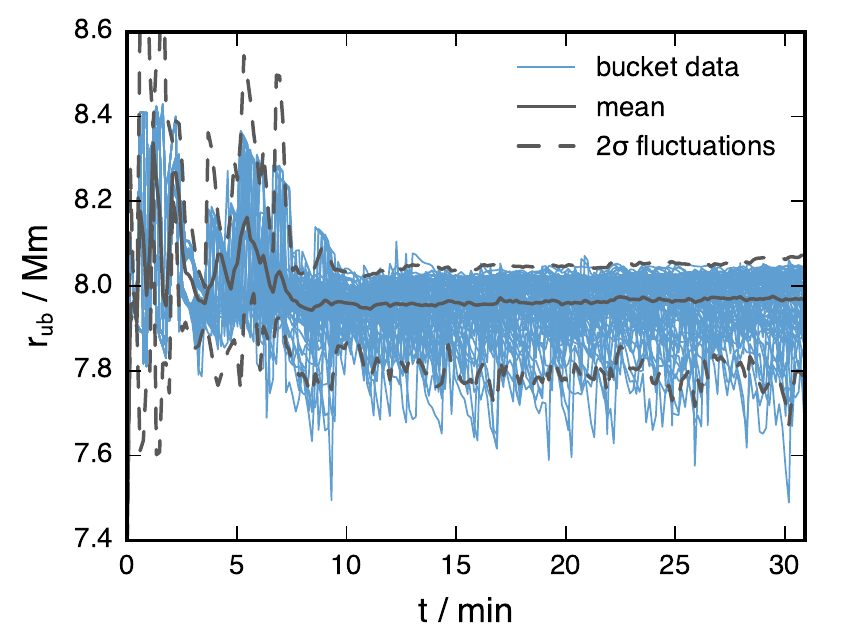}
 \includegraphics[width=0.49\linewidth, clip=true, trim=0mm 1mm 0mm
                  2mm]{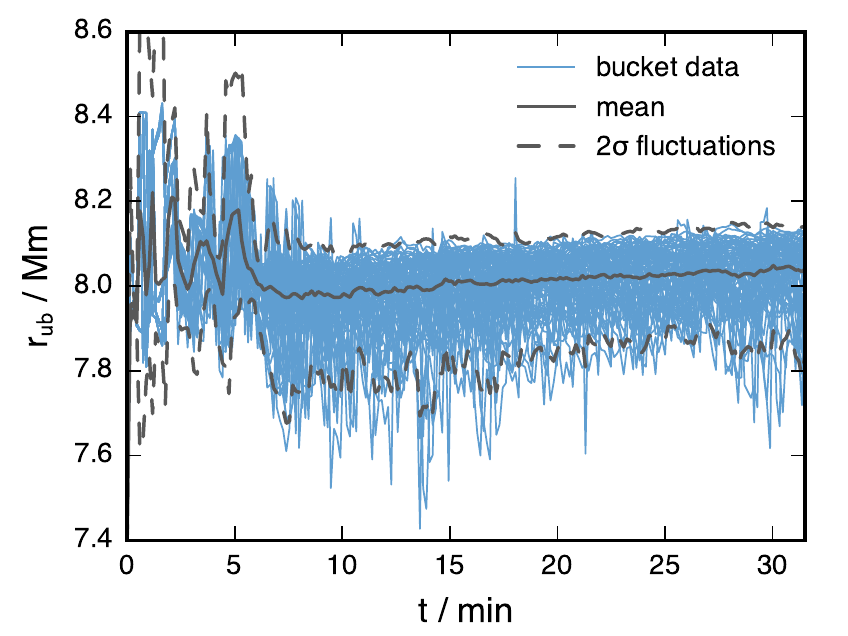}
\caption{Evolution of the radius $\rub$ of the upper boundary in the
\textsc{d1} (left panel) and \textsc{d8} (right panel) runs as defined by the
steepest gradient in the RMS tangential velocity (see
Figs.~\ref{fig:upper_boundary_vt} and \ref{fig:upper_boundary_dvtdr}). The full
sphere is split into 80 radial \emph{buckets} (see
Fig.~\ref{fig:bucket_map_r_ub} and Section~\ref{sec:1d-3d}). The standard
deviation $\sigma$ was computed separately for positive and negative
fluctuations. The large-amplitude oscillations in the first $\sim 10$ minutes
result from the initial transient (see Section~\ref{sec:PPM-setup}).}
\label{fig:upper_boundary_evolution}
\end{figure*}

In this section we will describe how convection in the 3D simulations presented
here differs from the 1D picture. The entropy gradient averaged spherically and
over a few overturning time scales is weakly positive in the upper half of the
convection zone. If interpreted as a 1D representation of the stratification it
would imply this upper part to be formally stable.
However, convection there is just as vigorous as that in the lower part
of the convection zone (see Figs.~\ref{fig:1536-vort}, \ref{fig:1536-vertvel},
and \ref{fig:vprof-22min}). This suggests that instead of the average
stratification the velocity field itself should be used to delineate the
boundaries of a 3D convection zone. Figure~\ref{fig:vprof-22min} shows that the
tangential velocity component $v_\perp$ reaches a local maximum both near the
bottom and near the top of the convection zone and then drops suddenly at the
convective boundary, where the steep entropy gradient forces the flow to turn
over. The flow velocity does not vanish completely because of the presence of
internal gravity waves and sound waves in the stable stratification.
We adopt the convective boundary as the location where the
decline in $v_\perp$ is the steepest.

This criterion can be applied to the $4\pi$ spherical average as well as to
different radial directions. This shows that in contrast to the 1D MLT picture,
the boundary of a 3D convection zone is not located on a perfect sphere. The
variation of the location of the convective boundary in different directions
translates into finite thickness of the boundary when spherical averages are
formed. To illustrate this, the data has been split into 80 space-filling
radial tetrahedra (which we call \emph{buckets}), each covering approximately
the same solid angle (see Fig.\,\ref{fig:bucket_map_r_ub}).
Figures~\ref{fig:upper_boundary_vt} and \ref{fig:upper_boundary_dvtdr} show
that there are indeed significant bucket-to-bucket fluctuations $\Delta \rub =
\rub - \langle \rub \rangle$ when the radius $\rub$ of the upper boundary is
defined by the steepest decline in $v_\perp$.

\afterpage{
\begin{figure*}
 \includegraphics[width=0.49\linewidth, clip=true, trim=0mm 2mm 0mm
                  2mm]{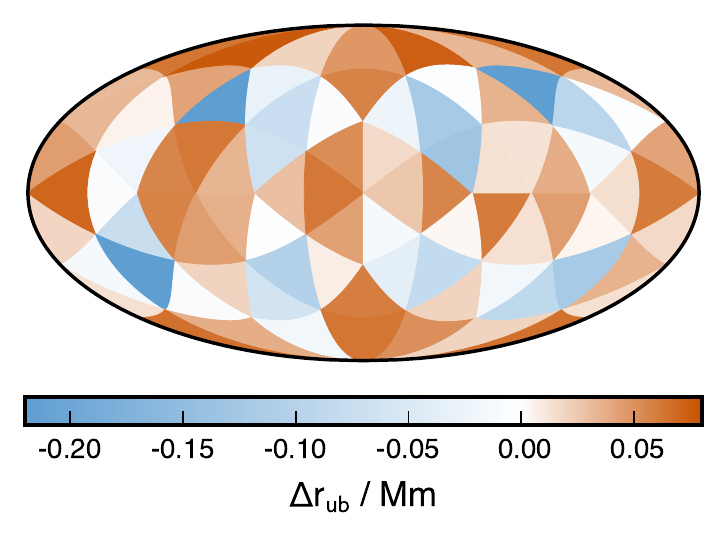}
 \includegraphics[width=0.49\linewidth, clip=true, trim=0mm 2mm 0mm
                  2mm]{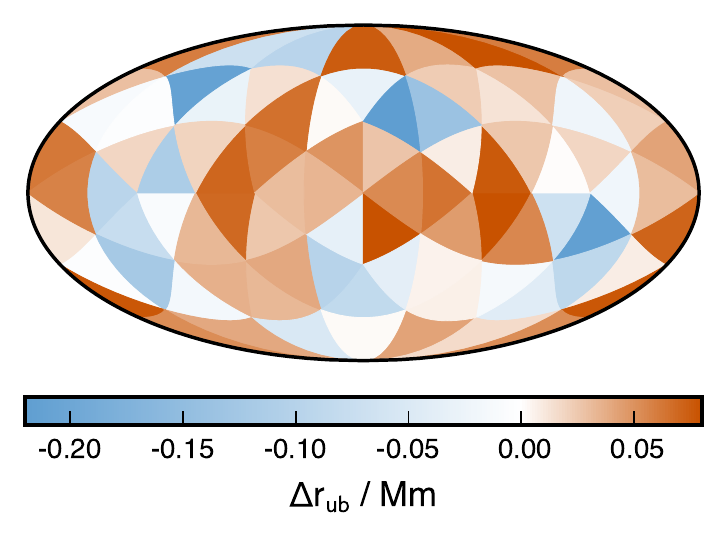}

\caption[]{Spatial distribution of the fluctuations $\Delta \rub =
\rub - \langle \rub \rangle$ in the radius of the upper
convective boundary (see Fig.~\ref{fig:upper_boundary_vt} and
Section~\ref{sec:1d-3d}) at $t = 18\,\mathrm{min}$ in the \textsc{d1} (left
panel) and \textsc{d8} (right panel) runs. Data from 80 individual
\emph{buckets} (triangular in cross-section, see Section~\ref{sec:1d-3d}) are
shown in an equal-area projection. The two large-scale circular concentrations
of mostly negative values and the lack of negative values on the rest of the
sphere in the \textsc{d1} data show the grid imprint, which is essentially
invisible in the \textsc{d8} data at this and later stages of the simulation
(this effect is clearer in the animated version of these
maps\protect\footnotemark).}

\label{fig:bucket_map_r_ub}
\end{figure*}

\footnotetext{See movies at the URLs given in Section~\ref{sec:conclusions}}
}

From the spatial point of view, the left panel of
Fig.~\ref{fig:bucket_map_r_ub} shows that the lowest-luminosity run \textsc{d1}
contains a distinct and persistent\footnote{Animated versions of the plots from
Fig.~\ref{fig:bucket_map_r_ub} show this effect clearly, see the movies at the
URLs given in Section~\ref{sec:conclusions}} pattern in the form of two
large-scale circular concentrations of mostly negative values and a lack of
negative values on the rest of the sphere. This effect results from the grid
imprint discussed in Sect.~\ref{sec:PPM-props}, but even in the
lowest-luminosity \textsc{d1} and \textsc{d2} runs it is weak enough not to
influence mass entrainment rates as shown in Sect.~\ref{sec:entrainment}. The
flow in the higher-luminosity runs is stronger and it quickly destroys the grid
imprint as illustrated in the right panel of Fig.~\ref{fig:bucket_map_r_ub} on
the example of the \textsc{d8} run, which has the second lowest luminosity
considered ($2.5 \times$ the luminosity of \textsc{d1}, see
Table~\ref{tab:run-info}).

The temporal evolution of the upper boundary is shown in
Fig.~\ref{fig:upper_boundary_evolution}. One can see that the boundary trembles
stochastically and has a thickness, as defined by the $2\sigma$ spatial
fluctuations shown in Fig.~\ref{fig:upper_boundary_evolution}, of $\sim
0.25$\,Mm ($0.17 H_\mathrm{P}$) and $\sim 0.3$\,Mm in the \textsc{d1} and
\textsc{d8} runs, respectively. Bucket data are not available for the rest of
our runs, but volume renderings\footnote{See movies at the URLs given in
Section~\ref{sec:conclusions}} suggest that the fluctuations at higher
luminosities are larger still as one would expect. As the material from above
the convection zone is entrained, the mean radius $\rub$ of the boundary
statistically increases. This effect is difficult to see in the left panel of
Fig.~\ref{fig:upper_boundary_evolution}, because the average velocity of the
boundary is small in the \textsc{d1} run, but the mean radius of the boundary
is clearly seen to increase with time in the right panel of
Fig.~\ref{fig:upper_boundary_evolution}, which shows the \textsc{d8} run with
$2.5\times$ the luminosity of \textsc{d1}. Measured values of
$\dot{r}_\mathrm{ub}$ are given in Table~\ref{tab:run-info}. The temporal
fluctuations in the mean boundary radius $\rub(t)$ around the linear trend are
about $\pm 7\times 10^{-3}$\,Mm in the \textsc{d1} and \textsc{d2} runs after
the initial transient has passed. When extrapolated to $t = 0$, the boundary
radius in the \textsc{d1} (\textsc{d2}) run is $7.99$\,Mm ($8.01$\,Mm). These
radii coincide within $\sim 0.07$\,Mm ($\sim 0.05 H_\mathrm{P}$; less than
$1/3$ of the spatial fluctuations in $\rub$, see above) with the radius
$r_\mathrm{SC} = 7.94$\,Mm of the formal Schwarzschild boundary in the
\textsc{PPMstar} setup.

\subsection{Mass entrainment at the upper convective boundary}
\label{sec:entrainment}

The entrained mass $M_\mathrm{e}(t)$ is defined in the present work to be the
total amount of the initially stable fluid $\mathcal{F}_1$ (see
Sect.~\ref{sec:PPM-setup}) present in the convection zone at time $t$,

\begin{equation}
M_\mathrm{e}(t) = 4\pi\int_{r_1(t)}^{r_2(t)}
\rho_1(r,\,t)\, r^2 \diff r,
\label{eq:entrained-mass}
\end{equation}
where $\rho_1(r,\,t)$ is the density of fluid $\mathcal{F}_1$ alone. The lower
integration limit is in practice taken to be $r_1(t) = 0$ (i.e. the
bottom of the simulation box), because the amount of fluid $\mathcal{F}_1$
getting below the convection zone is utterly negligible\footnote{This also
holds true for the high-luminosity runs, including \textsc{d10}.} (see
Fig.~\ref{fig:FV-profiles}). The upper integration limit $r_2(t)$
would ideally be taken equal to the radius $r_2=\rub(t)$ of the upper
boundary as defined in Sect.~\ref{sec:1d-3d} using the gradient of the
spherically-averaged RMS tangential velocity $v_\perp$. However, experience has
shown that the fractional volume of fluid $\mathcal{F}_1$ is so high at $r =
\rub$ as compared to the rest of the convection zone that the
resulting $M_\mathrm{e}(t)$ becomes very sensitive to small fluctuations in
$\rub(t)$. Therefore we use $r_2 = \rub(t) -
H_{v,\mathrm{ub}}(t)$, where the scale height $H_{v,\mathrm{ub}} = v_\perp
|\partial v_\perp / \partial r|^{-1}$ of $v_\perp$ is evaluated at $r = \rub(t)$.

\begin{figure*}
 \includegraphics[width=0.49\linewidth, clip=true, trim=0mm 3mm 0mm
                  2mm]{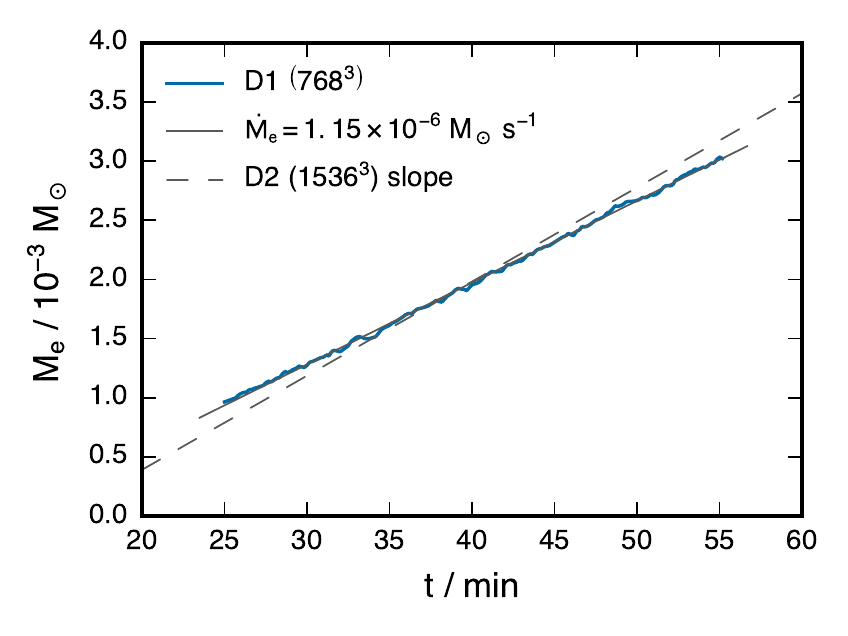}
 \includegraphics[width=0.49\linewidth, clip=true, trim=0mm 3mm 0mm
                  2mm]{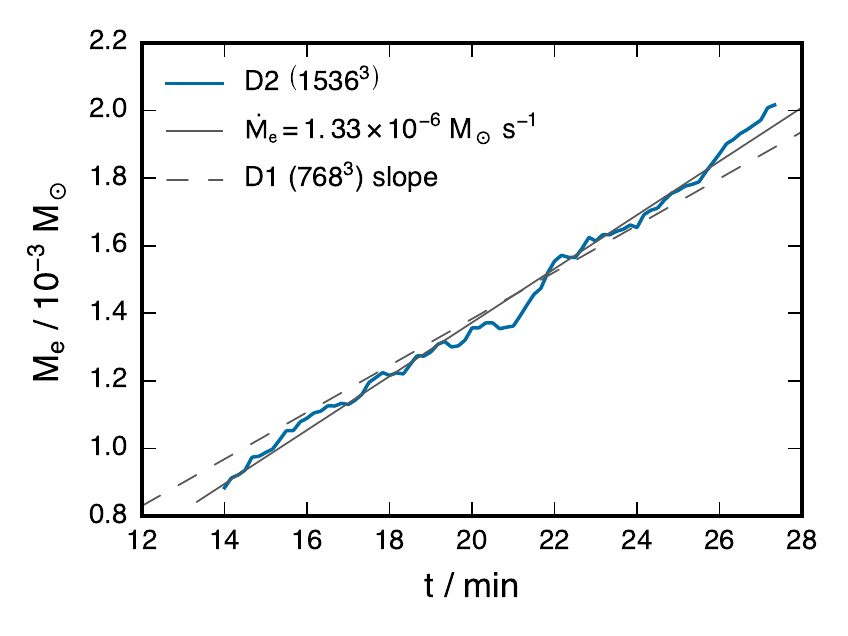}

 \caption{Entrained mass as a function of time for the $768^3$ run \textsc{d1}
(left panel) and the $1536^3$ run \textsc{d2} (right panel). The two
entrainment rates agree within 16 per cent. The dashed line in each
of the panels shows the best-fit slope from the other panel for comparison.
The values of $M_\mathrm{e}(0)$ have been adjusted for the dashed lines, because
different amounts of mass were entrained during the initial transients in the two
runs.}

 \label{fig:entrainment-rates}
\end{figure*}

The profiles of $M_\mathrm{e}(t)$ for runs \textsc{d1} and \textsc{d2} are
shown in Fig.~\ref{fig:entrainment-rates}. The initial stage of flow
development, which is influenced by the initial transient and grid imprints, is
not considered. The entrainment rate $\dot{M}_\mathrm{e}$ is calculated by
performing a least squares fit to the time series in $M_\mathrm{e}$, which was
constructed by applying the formula in Eq.~\ref{eq:entrained-mass} at each
available output dump (every $10$\,s of simulated time). The entrainment rate
obtained at a resolution of $1536^3$ (run \textsc{d2}) is $1.33 \times
10^{-6}\,M_\odot\,\mathrm{s}^{-1}$, which is only 16 per cent higher than the
entrainment rate of $1.15 \times 10^{-6}\,M_\odot\,\mathrm{s}^{-1}$ obtained at
$768^3$ (run \textsc{d1}).

The remaining runs listed in Table~\ref{tab:run-info} were analysed in the same
way. The profiles of $M_\mathrm{e}(t)$ and the associated best-fit entrainment
rates for these runs can be found in
Fig.~\ref{fig:entrainment-rates-other-runs} in
Appendix~\ref{sec:entrainment-rates-other-runs}. The entrainment in the two
highest-luminosity runs (\textsc{d9}, \textsc{d10}) was found to be so fast
that also a certain period of time at the end of the simulation had to be
removed from the entrainment analysis, because the flow started to become
influenced by the presence of the upper boundary of the simulation box well
before the end of the simulation.

The only parameter changed in the series of simulations \textsc{d1},
\textsc{d5}, \textsc{d6}, \textsc{d8}, \textsc{d9}, and \textsc{d10} is the
luminosity of the heat source, which spans almost two orders of magnitude. The
left panel of Fig.~\ref{fig:entrainment_rate_scalings} shows that the
entrainment rate $\dot{M}_\mathrm{e}$ is almost directly proportional to the
luminosity $L$. This scaling predicts an entrainment rate of $5.4 \times
10^{-7}\,M_\odot\,\mathrm{s}^{-1}$ for the luminosity of the actual stellar
model considered ($5.2 \times 10^{10}L_\odot$). 

Likewise, there is a tight relation between the entrainment rate and the shear
velocity at the upper boundary as shown in the right panel of
Fig.~\ref{fig:entrainment_rate_scalings}. The shear velocity is approximated by
the value of the RMS tangential velocity $v_\perp$ at the point where $v_\perp$
reaches a local maximum just below the upper boundary of the convection zone
(see Fig.~\ref{fig:vprof-22min}).

In Sect.~\ref{sec:1d-3d}, a potential concern was raised about the presence of
a persistent grid imprint in the \textsc{d1} and \textsc{d2} runs.
Figure~\ref{fig:entrainment_rate_scalings} shows that these two runs do not
deviate in any significant way from the scaling relations established by the
higher-luminosity runs. The grid imprint is apparently weak enough not to
influence the entrainment rates even at the low luminosity of the \textsc{d1}
and \textsc{d2} runs.

\subsection{Modelling convection as a diffusive process in spherically
symmetric models}
\label{sec:diffusion}

One of the important goals of simulating stellar convection in 3D hydrodynamic
codes such as \textsc{PPMstar} is to inform the treatments of convection and
convective boundary mixing in 1D stellar models.  In such spherically-symmetric
1D models, convection is most commonly approximated as a diffusive process.
Hence, the approach taken here to reconnect the mixing in the 3D simulation
with 1D simulations is to try to answer the question: \emph{what diffusion
coefficient profile would have been needed in 1D in order to reproduce the
spherically-averaged mixing in the 3D simulation?}

The diffusion equation was discretised to give
\begin{multline}
x_m(X_k^n - X_k^{n+1}) = \dfrac{X_{k}^{n+1} - X_{k-1}^{n+1}}{x_l}\Delta t~D_k + \\
\dfrac{X_{k}^{n+1} - X_{k+1}^{n+1}}{x_r}\Delta t~D_{k+1}~,
\label{eq:diffusion}
\end{multline}
where
\begin{eqnarray*}
x_r & = & r_{k+1} - r_k~, \\
x_l & = & r_k - r_{k-1}~, \\
x_m & = & \dfrac{1}{2} (r_{k+1} - r_{k-1})~,
\end{eqnarray*}
$X$ is the mass fraction of a particular component of the fluid, $D$ is the
diffusion coefficient, $r$ is the radius, $k$ and $n$ are spatial and temporal
indices, respectively, and $\Delta t$ is the time between temporal indices $n$
and $n+1$.

After having performed the 3D simulation in \textsc{PPMstar}, the mass fraction
$X(r,t)$ is known from spherical averages of the results, and of course the
time between spherically averaged profiles $\Delta t$ is also a known quantity.
Applying the boundary condition $D_1=0$ allows equation~\ref{eq:diffusion} to
be solved for $D(r)$ and, hence, provide an answer to the question posed above.

\subsubsection{General shape of the diffusion coefficient}
\label{sec:general-D}

The method described above has been used to calculate the 1D diffusion
coefficient profile across the whole depth of the O-burning convection zone
that represents the spherically and temporally averaged mixing of the 3D
hydrodynamic {\sc PPMstar} {\sc d2} simulation (see Table~\ref{tab:run-info})
between $t_1=620$~s and $t2=1140$~s.  The radial abundance profiles of the
upper fluid (as are shown in Fig.~\ref{fig:FV-profiles}) used for the initial
and final conditions of the diffusion problem are 400~s (on the order of a
convective turnover time scale) time averages centred at $t_1$ and $t_2$.  This
averaging smooths out some of the noise that is inherent when stochastic
fluctuations in the direction of inclination or azimuth appear on dynamical
time-scales. The inverted discretised diffusion equation is then solved for
$D(r)$ over the time step of $\Delta t=t_2-t_1$. The resulting diffusion
coefficient is shown as the brown line in Fig.~\ref{fig:simple-D-fit}.  When
the gradient of the abundance approaches 0 (essentially flat; see
Fig.~\ref{fig:FV-profiles}), there is not a unique solution to the inverse
diffusion problem, only a minimum diffusion coefficient that would completely
mix two neighboring zones when left for a time $\Delta t$, above which any
value would be a valid solution.  This is the origin of the noise seen in
Fig.~\ref{fig:simple-D-fit}: a loss of sensitivity of the method as $dX/dr$
approaches 0. In the noisy region, a piecewise-linear representation of a
downsampled diffusion coefficient was constructed, to which a cubic function
was fit, resulting in the solid black line in Fig.~\ref{fig:simple-D-fit}. This
represents a kind of by-eye fit to the lower envelope of the diffusion
coefficient profile in the very noisy region between radial coordinates of
about 5 and 6~Mm.

There are a few other interesting quantities plotted in
Fig.~\ref{fig:simple-D-fit} in addition to the diffusion coefficient derived as
described above. Firstly, the diffusion coefficient computed using MLT in a
representative MESA simulation is shown, which compares within an order of
magnitude, underestimating in the middle of the convection zone (where the
method used to derive $D$ loses sensitivity) and overestimating near the
convective boundary. The light blue dashed line marked with circular glyphs is
a diffusion coefficient given by the same formula as the one used in MLT
($D=\frac{1}{3}v_\mathrm{PPM}\alpha H_P$) but $v_\mathrm{PPM}$ is the average
radial velocity of the convective fluid from the 3D {\sc PPMstar} simulation.
It is within a factor of $\sim2$ of the value predicted using MLT.
When one considers that the luminosity of the {\sc PPMstar} simulation is
generally within a factor $\sim 2$ of the MESA model (see
Fig.~\ref{fig:mesa-props}) together with the scaling of velocity with
luminosity (Fig.~\ref{fig:radial_velocity_vs_luminosity}), the {\sc PPMstar} and MLT velocities
agree within a factor of $\sim 1.6$ inside the convection zone.  Interestingly,
\citet{herwig:06a} also found in their 2D simulations that the MLT velocities
were too low by a factor of about 2--3 (their Fig.~24). Such a difference
between 3D hydro and 1D MLT velocities can be due to neglect of kinetic energy
flux in MLT or the different geometric and averaging approximations made in
various flavours of MLT \citep[see, e.g.,][]{Tassoul1990,Salaris2008}.

In any case, using $\frac{1}{3}v_\mathrm{PPM}\alpha H_P$ overestimates $D$ in
the upper and lower regions of the convective layer, too.  The blue starred
line looks somewhat different.  This line assumes instead that the mixing
length is the minimum of $\alpha H_P$ and the distance to the upper convective
boundary.  This reproduces better the shape of the derived D as it falls off in
the upper half of the convection zone. The reduction of such a mixing length as
the flow approaches the upper convective boundary was also suggested by
\citet{eggleton:72} and is analogous to the decrease of the horizontal scale of
convection cells over a similar domain, as has been shown by \citet[][their
Fig.~5]{Porter:2000eo}.

\begin{figure*}
 \includegraphics[width=0.49\linewidth, clip=true, trim=0mm 3mm 0mm
                  2mm]{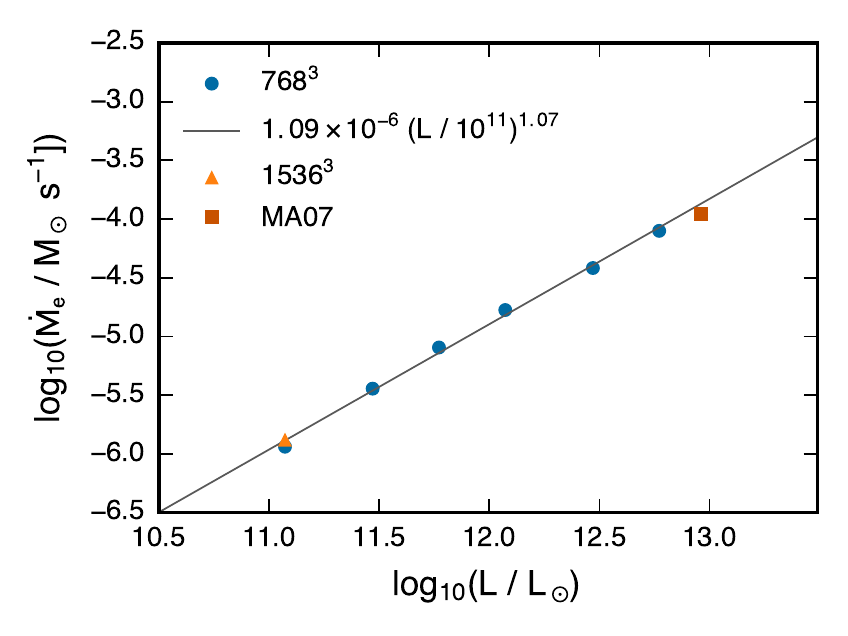}
 \includegraphics[width=0.49\linewidth, clip=true, trim=0mm 3mm 0mm
                  2mm]{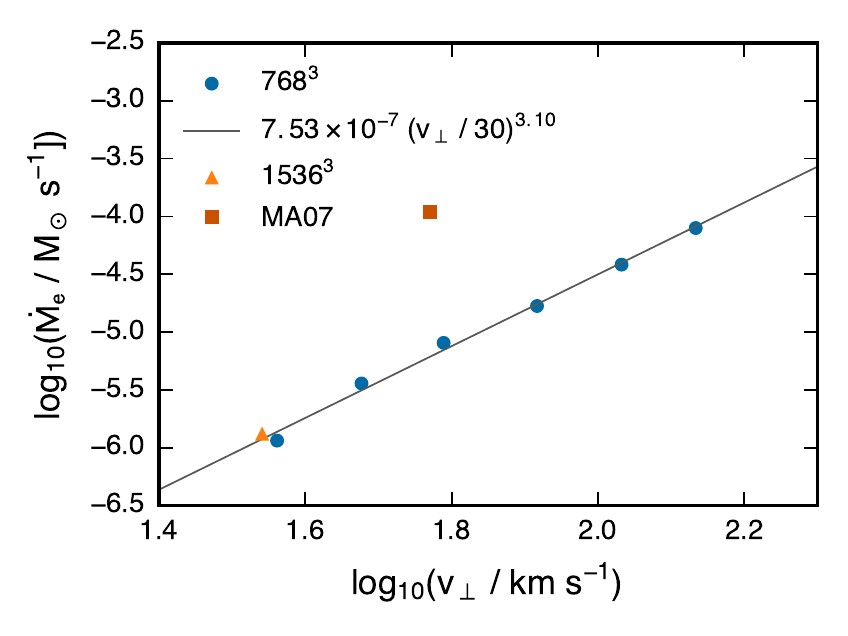}

\caption{Dependence of the entrainment rate $\dot{M}_\mathrm{e}$ on the
luminosity $L$ (left panel) and on the shear velocity $v_\perp$ (right panel).
The shear velocity $v_\perp$ is approximated by the value of the RMS tangential
velocity $v_\perp$ at the point where $v_\perp$ reaches a local maximum just
below the upper boundary of the convection zone (see
Fig.~\ref{fig:vprof-22min}). The RMS velocity profiles were averaged over
$10$ data dumps (spanning 100~s of simulation time) centred in the middle of
the time interval, in which the mass entrainment rate was measured for the
given run. The fitting relations, which assume the same units as the axes of
the plots, were constructed using the $768^3$ runs only. The $1536^3$ run
\textsc{d2} and the 3D simulation of \citet{Meakin2007a} are shown for
comparison.}

\label{fig:entrainment_rate_scalings}
\end{figure*}

\subsubsection{Convective boundary mixing}
\label{sec:cbm}

The results of the determination of a radial diffusion coefficient profile are
shown in Fig.~\ref{fig:DDDDD_boundary} for the upper convective boundary of the
{\sc d1} and {\sc d2} simulations.  The vertical dotted lines mark the upper
boundary of the convection zone between the convectively unstable (on the left
of the boundary) and stable (on the right) fluids. The two lines represent
different definitions of the convective boundary.  B is the radius at which the
entropy gradient becomes positive in the initial stratification of the 3D
simulations and C is where the radial gradient of the tangential component of
the fluid velocity is steepest in the 3D simulation (see
Section~\ref{sec:1d-3d}) after 40 (10) minutes of simulated time for the {\sc
d1} ({\sc d2}) run. This is where the convective boundary is in 3D. The
underlying MESA model has been aligned so that its Schwarzschild boundary lies
exactly on top of B, so that a better comparison can be made between the 3D and
1D simulation results.

The abundance of the upper (initially stable) fluid is shown as a function of
radius at two different times\footnote{The actual subscript of $X$ in
Fig.~\ref{fig:DDDDD_boundary} denotes the dump number of the profile
quantities.  In this simulation, the dump interval was 10 seconds of simulated
time.} by $X_1$ (brown line corresponding to time $t_1$) and $X_2$ (grey line
corresponding to time $t_2$). The abundance of the upper fluid changes over the
domain of the figure as the fluid from the overlying stable layer is entrained
into the convection zone. The solid light blue curve shows the diffusion
coefficient $D_\mathrm{MLT}=\frac{1}{3}v_\mathrm{MLT}\alpha H_P$ (with
$\alpha=1.6$) used in the underlying MESA model using mixing length theory to
estimate the convective velocities. The diffusion coefficient resulting from
our calculation is shown as a solid black curve. The result was verified by
solving the diffusion equation over the {\sc d2} problem domain using $X_{62}$
as the initial abundance profile, $\Delta t = t_2 - t_1$ as the time step and
our spatially-varying diffusion coefficient. The abundance profile $X_{130}$ at
$t_1 + \Delta t$ was obtained exactly. The method used here can only
distinguish mixing in which some of the upper and lower fluids have been
exchanged. Because of this, we are not able to constrain the mixing above
roughly $8.25\times10^8$~cm (for {\sc d1}; $8.15\times10^8$~cm for {\sc d2})
using the present methods with the results from the simulations presented in
this work. Combining the exponentially decaying mixing model of
\citet{freytag:96} and the mixing length modification described in
Section~\ref{sec:general-D}, the shape of the diffusion coefficient can be very
well reproduced in a 1D model. This is plotted as $D_\mathrm{RCMD}$ and will be
explained in greater detail in Section~\ref{sec:3d1d}. The difference between
the two panels in Fig.~\ref{fig:DDDDD_boundary} is the numerical resolution of
the {\sc PPMstar} simulations: $768^3$ in the top panel and $1536^3$ in the
bottom panel. The exponential decay of $D_\mathrm{RCMD}$ across the convective
boundary in both panels is best fit with an e-folding length of
$0.03\frac{H_P}{2}$ (i.e., $f_\mathrm{CBM}=0.03$ in Eq.~\ref{eq:expD}). It is a
very encouraging result that the shape and scale height of the derived
diffusion coefficient changes so little upon refinement of the computational
grid. In Appendix~\ref{sec:f-determination}, the same plots are shown with
different values of $f_\mathrm{CBM}$, illustrating the suitability of
$f_\mathrm{CBM}=0.03$ to this particular convective boundary.

\begin{figure}
\includegraphics{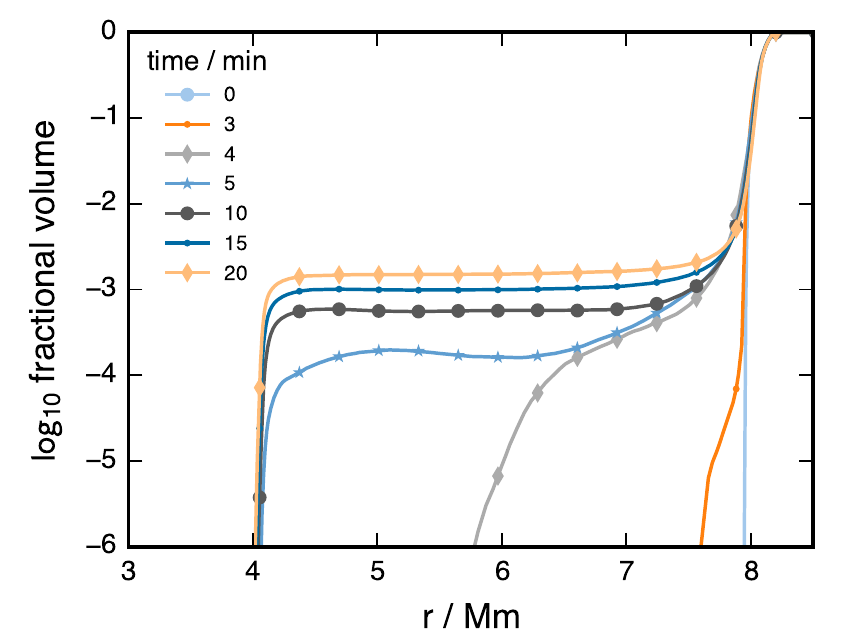}

\caption{Spherical averages of the fractional volumes of the upper (initially
stable) fluid as a function of radial coordinate at different times over the
duration of the {\sc d2} simulation.}

\label{fig:FV-profiles}
\end{figure}

\subsection{A mixing prescription for 1D codes}
\label{sec:3d1d}

In Fig.~\ref{fig:DDDDD_boundary} a recommended diffusion coefficient profile is
drawn that, based on our {\sc PPMstar} simulations, is a better representation
of the space and time averaged mixing during O shell burning than using MLT
with the Schwarzschild or Ledoux criteria. It is a combination of two simple
considerations that one can easily adopt in a 1D stellar evolution code.
Firstly, instead of using simply $\ell=\alpha H_P$ for the mixing length, where
$\alpha$ is a constant of order unity, using the minimum of $\ell$ and the
distance to the upper convective boundary as the mixing length, as suggested by
\citet{eggleton:72}, much better reproduces the fall-off of the diffusion
coefficient inside the convection zone approaching the convective boundary (see
Fig.~\ref{fig:simple-D-fit} and Section~\ref{sec:general-D}).  The recommended
diffusion coefficient is then written as
\begin{equation}
D_\mathrm{RCMD} = v_\mathrm{MLT}\times\min(\alpha H_P,|r-r_\mathrm{SC}|)\;,
\label{eq:Drcmd}
\end{equation}
where $r_\mathrm{SC}$ is the radius of the Schwarzschild convective boundary.
$D_\mathrm{RCMD}$ contains a constant factor of 3, which has cancelled
the factor of one third that usualy appears in Eq.~\ref{eq:Drcmd}. This factor
accounts for the remaining discrepancy between the RMS radial velocities from
the 3D hydrodynamic simulation and those predicted using MLT (see
Section~\ref{sec:general-D}) after the velocities have been re-scaled by the
relation shown in Fig.~\ref{fig:radial_velocity_vs_luminosity} for better agreement between the
driving luminosities in the {\sc PPMstar} and MESA simulations.  Secondly,
applying the exponentially decaying diffusive convective boundary mixing model
of \cite{freytag:96} in combination with the modified mixing length seems to
reproduce the shape of the derived diffusion coefficient rather well. In
Fig.~\ref{fig:DDDDD_boundary} this has been plotted, and is formulated
for the upper convective boundary as
\begin{equation}
D(r)=D(r_0)\times\exp\left\{-\dfrac{2|r-r_0|}{f_\mathrm{CBM}H_{P}(r_0)}\right\}
\label{eq:expD}
\end{equation}
where $r_0$ is the radial coordinate $r_\mathrm{SC}-f_\mathrm{CBM}H_P^\mathrm{SC}$
(i.e. inside the convection zone, with
$H_P^\mathrm{SC}$ the scale height of pressure at the Schwarzschild boundary. Lastly,
$f_\mathrm{CBM}$ is the parameter that sets the e-folding length of the
exponentially-decaying diffusion coefficient---which is $f_\mathrm{CBM}/2$ in
units of the presure scale height $H_P$---outward from the radius $r_0$. In
Fig.~\ref{fig:DDDDD_boundary} a value of $f_\mathrm{CBM}=0.03$ was used.  Note
that this mixing model is already implemented in the MESA stellar evolution
code, and the recommendation made in this section would be implemented by
setting the parameters ${\tt overshoot\_f0\_above\_burn\_z\_shell} = 0.03$ and
${\tt overshoot\_f\_above\_burn\_z\_shell} = 0.03$ with an additional minor
modification to multiply the diffusion coefficient by the factor 3.

\subsection{Temperature fluctuations and their feedback on the energy
generation rate}

Our use of a static heat source instead of a nuclear network raises the question
of what effect the convection-induced temperature fluctuations could have on the
energy generation rate in a real star. The temperature sensitivity of
$^{16}\mathrm{O}+\!^{16}\mathrm{O}$ reactions,
\begin{equation}
\nu_\mathrm{OO} = \frac{\partial \ln \epsilon_\mathrm{OO}}{\partial \ln T},
\end{equation}
with the reaction rate $\epsilon_\mathrm{oo}$ given by Eq.~18.75 of
\citet{kippenhahn:12} is
\begin{equation}
\nu_\mathrm{OO} = -\frac{2}{3} + 45.31 T_9^{-1/3} - 0.419 T_9^{2/3} - 0.593
T_9^{4/3} + 0.0206 T_9^2,
\label{eq:nu_OO}
\end{equation}
where $T_9 = T/(10^9\,\mathrm{K})$; the temperature sensitivity of the electron
screening factor has been neglected. Eq.~\ref{eq:nu_OO} gives
\mbox{$\nu_\mathrm{OO} = 32$} for $T_9 = 2.2$, which is the typical temperature
at the bottom of the O shell convection zone. The relative RMS temperature
fluctuations reach $T^\prime_\mathrm{rms} = 5\times 10^{-5}$ in the heating
layer in the \textsc{d1} run. The expected relative RMS fluctuations in the
energy generation rate are then $\epsilon_\mathrm{OO}^\prime =
\nu_\mathrm{OO}T^\prime_\mathrm{rms} \approx 2\times 10^{-3}$. With such a small
value of $\epsilon_\mathrm{OO}^\prime$, it seems safe to neglect the
fluctuations in the energy generation rate even after having allowed for some
spread around the RMS values.

\section{Comparison to other works}
\label{sec:discussion}

As shown in Fig.~\ref{fig:radial_velocity_vs_luminosity}, the
luminosity-velocity scaling established by our set of simulations agrees within
$\sim 0.15$\,dex with the maximum radial velocity reached in the 3D simulation
of O shell convection by \citet[][MA07]{Meakin2007a} (see their Fig.~6). This
agreement is surprisingly good given the significant differences in physics and
geometry between the simulations. However, a more detailed comparison of our
Fig.~\ref{fig:vprof-22min} with Fig.~6 of MA07 reveals that in the work of
MA07:
\begin{enumerate}[1.]
\item the decrease in the flow speed towards the top of the convection zone
is much more pronounced and
\item the local maxima in the tangential velocity at
the convective boundaries are narrower.
\end{enumerate}
The velocity field in the simulation of MA07 may be influenced by their
constraining the flow to a narrow ($30\times 30$\,deg$^2$) wedge whereas
Figs.~\ref{fig:1536-FVs} and \ref{fig:1536-vertvel} show that the flow develops
large-scale structures if it is allowed to. Another contribution to the
decrease of the flow speed with radius could be provided by neutrino cooling,
which is neglected in this work, but included in the work of MA07. The velocity
field at the bottom of the convection zone is likely influenced by the
structure of the heating source, which drives the flow. MA07 use a nuclear
reaction network comprising 25 nuclei to compute the heating rate.
Figure~\ref{fig:setup_energies} shows that the artificial heat source used in
our work is more extended and centred at a slightly larger radius compared to
the energy source that would be provided by the actual nuclear
reactions.\footnote{Although the detailed nuclear network used in the MESA
simulations shown in Fig.~\ref{fig:setup_energies} likely differs from that
used by MA07, one would expect qualitatively the same behaviour.} Preliminary
results from a \textsc{PPMstar} run with a simple nuclear network indeed show a
flatter $v_\mathrm{r}(r)$ profile with narrower maxima in $v_\perp(r)$ at the
convective boundaries.

\begin{figure}
\includegraphics[width=\linewidth]{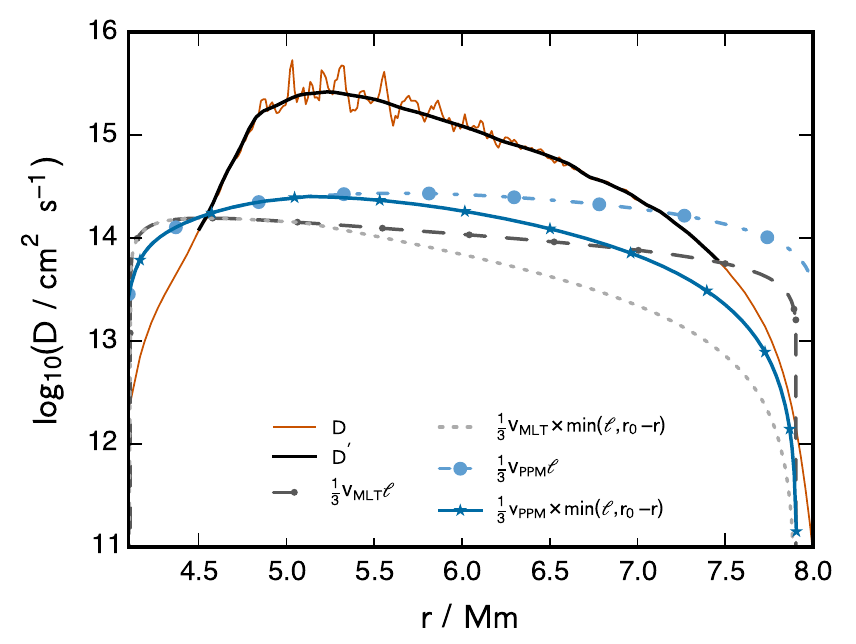}

\caption{Time-averaged radial diffusion coefficient profile calculated from the
spherically-averaged abundance profiles by the method described in
Section~\ref{sec:diffusion} (brown solid line; black solid line is a fit to the
noisy region). The convective velocities computing using MLT agree with the
spherically-averaged 3D velocities to within about a factor of 2 inside the
convection zone but are too large in the vicinity of the convective boundary,
resulting in an overestimation of the diffusion coefficient there. Limiting the
mixing length to the distance from the convective boundary reproduces the
fall-off of the diffusion coefficient inside the convection zone approaching
the boundary that is seen in the spherically averaged 3D simulation results.}

\label{fig:simple-D-fit}
\end{figure}

The flow in the 3D simulations of a similar O shell by \citet{Kuhlen2003}
reaches an RMS velocity of $49$\,km\,s$^{-1}$. Based on the velocity scaling
shown in Fig.~\ref{fig:radial_velocity_vs_luminosity} and the RMS velocity of $\sim
40$\,km\,s$^{-1}$ observed in the \textsc{d2} run (see
Sect.~\ref{sec:velocity-profiles}), one would expect and RMS velocity of $\sim
60$\,km\,s$^{-1}$ at the luminosity of $3.9 \times 10^{11}\,L_\odot$ used by
\citet{Kuhlen2003}. These authors simulate the whole sphere, although it is
unclear how the heating is modelled. They obtain a rather flat velocity
profile.  Unfortunately, the information they give about the velocity
components does not allow a detailed comparison.

The left panel of Fig.~\ref{fig:entrainment_rate_scalings} illustrates that the
3D simulation of MA07 follows the luminosity-entrainment-rate scaling
established by our simulations surprisingly well. However, the right panel of
the same figure shows that the entrainment rate of $1.1 \times
10^{-4}\,M_\odot$\,s$^{-1}$ is measured by MA07 at a tangential velocity $\sim
2.5$ times lower compared to our velocity-entrainment-rate scaling. The
stability of the upper convective boundary, as measured by the profile of the
squared buoyancy frequency $N^2(r)$, in the simulation of MA07 is very similar
to our set-up (compare their Fig.~2 with our Fig.~\ref{fig:N2_MESA_vs_PPM}).
However, as discussed in Sect.~\ref{sec:PPM-props}, entrainment in our
simulations is mostly observed at places where two neighbouring large-scale
convective cells meet and draw the partially-mixed material from the convective
boundary to the interior of the convection zone. There are only a few such
places in our simulations, but there could be many more (as extrapolated to the
full sphere) in the simulation of MA07, in which the artificial lateral
boundaries of their simulation box constrain the largest spatial scales in the
velocity field (discussed above). This effect could compensate the velocity
disadvantage mentioned above.

\citet{Spruit15} argues that the entrainment rate is limited by the buoyancy
arising from the relatively lower mean molecular weight of the entrained
material. His simple model predicts a linear dependence of the entrainment rate
on the convective luminosity, which is close to the relation measured in this
work (see the left panel of Fig.~\ref{fig:entrainment_rate_scalings}). In the
case of the entrainment of a fluid with a mean molecular weight $\mu_1$ into a
convective fluid with a mean molecular weight $\mu_2 > \mu_1$, Eq.~10 of
\citet{Spruit15} becomes:
\begin{equation}
\mu_x = \frac{\partial \ln \mu}{\partial x}\bigg|_{P, T} = -\mu\frac{\mu_2 - \mu_1}{\mu_2\mu1}
\end{equation}
where $x$ is the mass fraction of the lighter fluid, and $\mu$ the mean
molecular weight of the mixture. The equivalent of Spruit's Eq.~12 is then:
\begin{equation}
\dot{M}_\mathrm{i} = \frac{2}{5} \alpha_\mathrm{i} \frac{\mu_2\mu_1}{\mu_2 - \mu_1} \frac{L}{\mathcal{R}T},
\label{eq:mdot_i}
\end{equation}
where the subscript $\mathrm{i}$ refers to ``ingestion'' (which is called
``entrainment'' in this work), $\alpha_\mathrm{i}$ is a dimensionless
efficiency parameter, and $\mathcal{R} = 8.31 \times
10^7$\,erg\,g\,$^{-1}$\,K$^{-1}$ the gas constant. In the O shell simulations
presented in this work, $\mu_1 = 1.802$ and $\mu_2 = 1.848$, so
Eq.~\ref{eq:mdot_i} becomes
\begin{equation}
\dot{M}_\mathrm{i} = 29.0\, \alpha_\mathrm{i} \frac{L}{\mathcal{R}T},
\label{eq:mdot_i_2}
\end{equation}
which contains a numerical factor $12$ times larger than that in Eq.~12 of
\citet{Spruit15}, because the two fluids in the O shell problem are much more
similar in terms of $\mu$ than the helium and helium-carbon fluids in the
core-helium-burning problem investigated by \citet{Spruit15}. The temperature
at the top of the O-shell convection zone is $T = 1.4 \times 10^9$\,K. At the
luminosity of $L = 4.53 \times 10^{44}$\,erg\,s$^{-1}$ (runs {\sc d1} and {\sc
d2}), Eq.~\ref{eq:mdot_i_2} gives $\dot{M}_\mathrm{i} = 5.7 \times 10^{-5}
\alpha_\mathrm{i}$\,$M_\odot$\,s$^{-1}$ and the entrainment rate measured in
{\sc d2} corresponds to the efficiency parameter $\alpha = 2.3 \times 10^{-2}$.
For comparison, \citet{Spruit15} estimates $\alpha = 0.1$ for the ingestion of
helium into a helium-burning core.

\section{Summary and conclusions}
\label{sec:conclusions}
In this work, 3D hydrodynamic simulations representing the first O-burning
shell following core O extinction in a $25~M_\odot$ star were performed using
the {\sc PPMstar} code. The initial setup of the stratification was constructed
using three piecewise polytropes calibrated to match the O shell-burning
structure of a $25~M_\odot$ MESA model with initially solar metallicity
($Z=0.02$).  The luminosity from O burning was approximated using a constant
volume heating rate that also accounted for energy losses due to thermal
neutrino production and emission.  The simulations were performed on $768^3$
and $1536^3$ computational grids, including a heating series at $768^3$ in
which several simulations were performed, each with a heating rate scaled from
the original value by a constant factor.

At the representative heating rate of $1.18\times10^{11}~\lsun$, the $1536^3$
run {\sc d2} entrained material into the convection zone at the top boundary at
a rate of $1.33\times10^{-6}~\msun$~s$^{-1}$, while the $768^3$ run {\sc d1}
did so at a rate of $1.15\times10^{-6}~\msun$~s$^{-1}$. This is an increase of
only 16 per cent with twice the numerical grid resolution in each spatial
dimension.  The upper convective boundary is best defined at the radius with
the steepest radial gradient in the tangential velocity and the $2\sigma$
fluctuations in the radius of the convective boundary over the full $4\pi$
domain are about $0.2\times10^8$~cm. Larger fluctuations are present for higher
driving luminosities. The entrainment rate at the upper convective boundary
scales almost linearly with the driving luminosity of the convection and
approximately with the cube of the peak RMS shear velocity at the upper
convective boundary.  The heating series also reproduces the expected scaling
of velocities with the cube root of the driving luminosity.

\begin{figure*}
 \includegraphics[width=\linewidth, clip=true, trim=0mm 5mm 0mm
                  2mm]{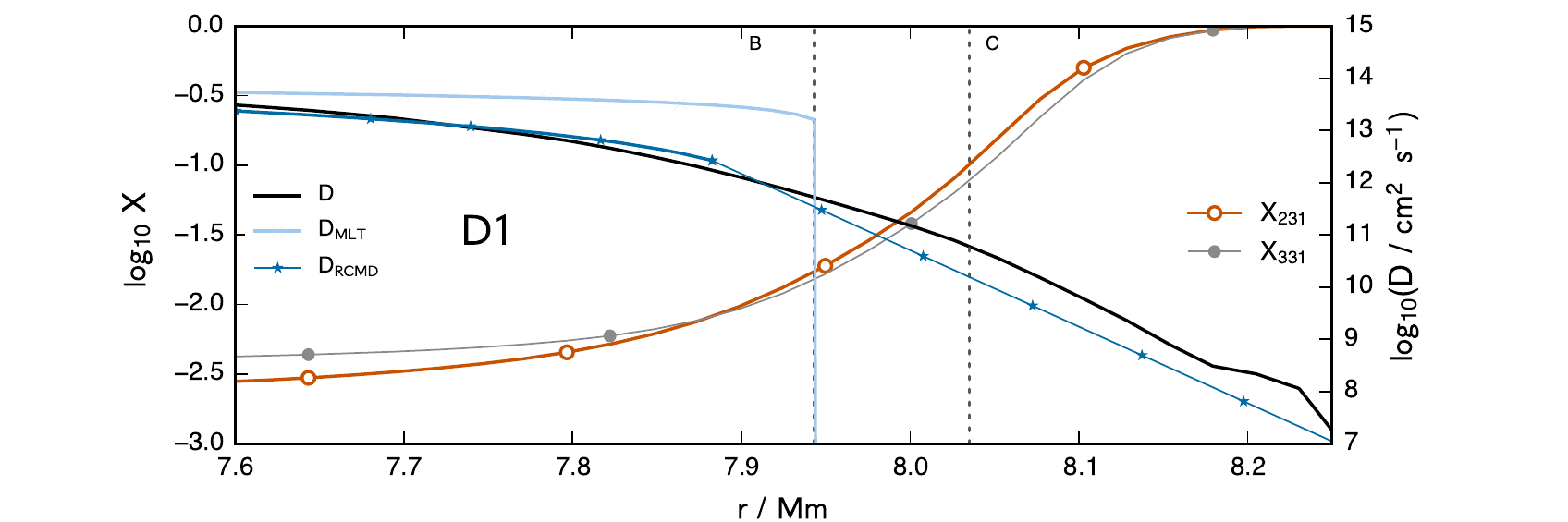}
 \includegraphics[width=\linewidth, clip=true, trim=0mm 0mm 0mm
                  0mm]{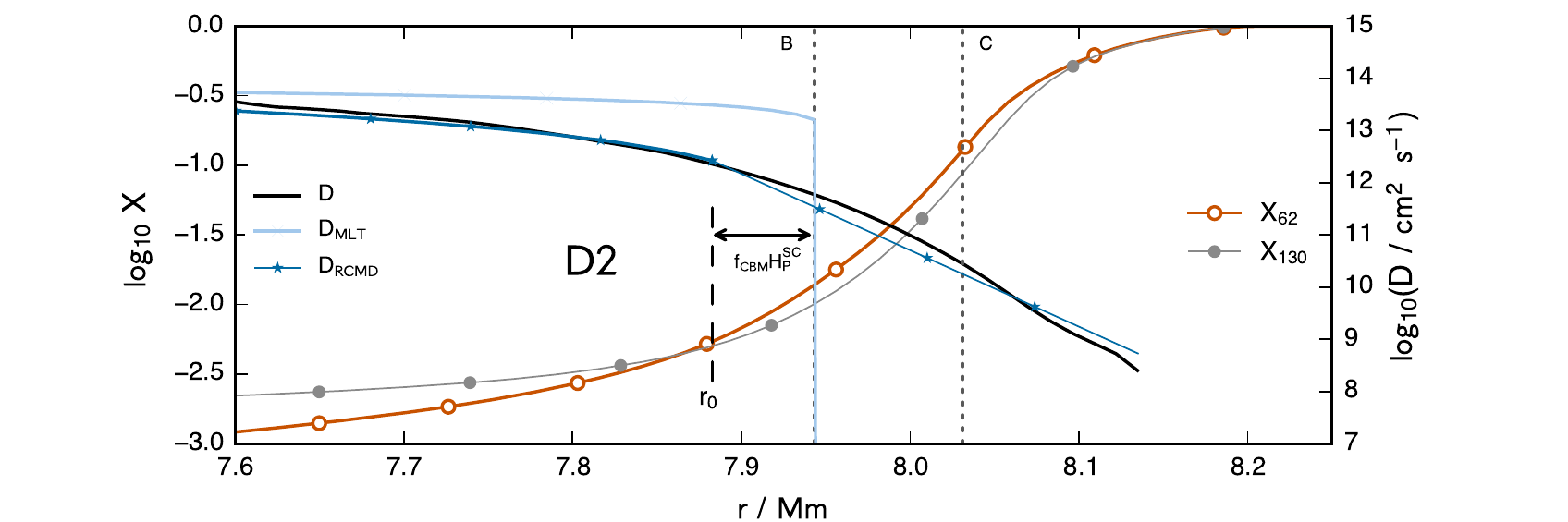}

\caption{Results of the 3D--1D diffusion analysis at the upper convective
boundary of the {\sc d1} ($768^3$ grid) and {\sc d2} ($1536^3$ grid)
simulations (see Table~\ref{tab:run-info}). The vertical dotted lines represent
the upper boundary of the convection zone. B is where the entropy gradient
becomes positive in our \textsc{PPMstar} setup (equivalent to the Schwarzschild
criterion); C is where the radial gradient of the tangential component of the
fluid velocity is steepest after 46.7 (16.0) minutes of simulated time for
simulation {\sc d1} ({\sc d2}). We also give the MESA model upon which these
simulations were based; it has been aligned so that the convective boundary
according to the Schwarzschild criterion is located at B. $X$ is the
spherically-averaged mass fraction of the overlying fluid and is plotted at a
simulated time indicated by the subscript in tens of seconds.  $D_\mathrm{MLT}
= \frac{1}{3}v_\mathrm{MLT}\alpha H_P$ is the diffusion coefficient computed in
the framework of mixing length theory with $\alpha=1.6$.  $D$ (solid black
line) is the derived diffusion coefficient that gives the same net mixing as
the 3D hydrodynamic simulation when its output is spherically averaged.
$D_\mathrm{RCMD}$ is the recommended diffusion coefficient to use in a 1D code
given by $D_\mathrm{RCMD}=v_\mathrm{MLT}\times\mathrm{min}(\alpha
H_P,|r-r_\mathrm{SC}|)$, where $r_\mathrm{SC}$ is the radial coordinate of the Schwarzschild
boundary at B, as described in Section~\ref{sec:3d1d} of the text, with an
exponentially decaying convective boundary mixing from radius
$r_0=r_\mathrm{SC}-f_\mathrm{CBM}H_P$ with $f_\mathrm{CBM}=0.03$, as formulated by
\citet[][see Section~\ref{sec:3d1d}]{freytag:96}.}

\label{fig:DDDDD_boundary}
\end{figure*}

In order to inform approximate mixing models for 1D stellar models, the 3D
simulation data was folded down to give the average values of quantities on a
sphere. Averaging the resulting radial profiles over approximately one
convective turnover timescale then gives spherically symmetric profiles that
may be comparable to those obtained from 1D stellar evolution codes that take
time steps much longer than a dynamical time. Diffusion coefficient profiles
that accurately represent the mixing of the radially-averaged 3D simulation
data are comparable with those computed using MLT. However, MLT overestimates
the diffusion coefficients near the convective boundary and likely
underestimates them inside the convection zone. This can be corrected by
limiting the mixing length to the distance to the convective boundary, as has
been suggested by \citet{eggleton:72}, and multiplying the diffusion
coefficnent by a factor of 3. Combining this simple modification with the
exponentially-decaying diffusion model of \citet{freytag:96} at and across the
convective boundary gives a mixing model for O shell burning in 1D codes that
reproduces the spherically-averaged abundance profile evolution. For the case
of O shell burning the parameter $f_\mathrm{CBM}$ in Eq.~\ref{eq:expD} at the
upper convective boundary that best reproduces the spherically-averaged mixing
in the 3D hydrodynamic simulation is $f_\mathrm{CBM}=0.03$.

High resolution simulations of stellar convection with an idealised approach to
microphysics appear to be a very useful technique with which to study stellar
convection in the O shell-burning episodes of massive stars. As others have
found, a key challenge is the interpretation of 3D simulation data in the
context of spherically symmetric 1D stellar models. This must of course involve
both the geometric averaging of quantities that fluctuate over $4\pi$ spheres
and the temporal averaging of quantities over long enough timescales to be
meaningful in the context of the assumptions that are made for hydrostatic 1D
models. The analysis presented in Section~\ref{sec:diffusion} is a somewhat
pragmatic step in a useful direction, but it is not intended to be the
predictive model that is so highly sought-after in the stellar physics
community.

\vspace{10pt}

Movies of the simulations performed in this work are available at
\url{http://www.lcse.umn.edu} and
\url{http://csa.phys.uvic.ca/research/stellar-hydrodynamics/movies/o-shell-convection-movies}

\section*{Acknowledgments}

SJ is a fellow of the Alexander von Humboldt Foundation and
acknowledges support from the Klaus Tschira Stiftung (KTS). RA, a CITA
national fellow, acknowledges support from the Canadian Institute for
Theoretical Astrophysics.  This work has been supported by NSF grant
PHY-1430152 (JINA Center for the Evolution of the Elements). The 3-D
hydrodynamical simulations reported here were carried out in part on
the Compute Canada WestGrid Orcinus cluster at UBC, Canada and in part
on the Blue Waters machine at NCSA under the support of NSF PRAC
awards 0832618 and 1440025 and CDS\&E award 1413548 to the University
of Minnesota. FH acknowledges support from a NSERC Discovery
grant. The development of the PPM code used in this work has also been
supported by contracts with the Los Alamos National Laboratory and the
Sandia National Laboratory through the University of Minnesota. FH
would like to thank Roman Baranowski from UBC/WestGrid for his kind
support and SJ and FH thank Michael Bennett for his preliminary work
on the diffusion analysis presented in this paper.  We would
  like to thank Bernhard M\"uller for several helpful comments.

\bibliography{papers2}  

\appendix \section{Mass entrainment in the runs \textsc{d5}, \textsc{d6},
\textsc{d8}, \textsc{d9}, and \textsc{d10}}
\label{sec:entrainment-rates-other-runs}

\begin{figure*}
 \includegraphics[width=0.49\linewidth, clip=true, trim = 0mm 8mm 0mm
0mm]{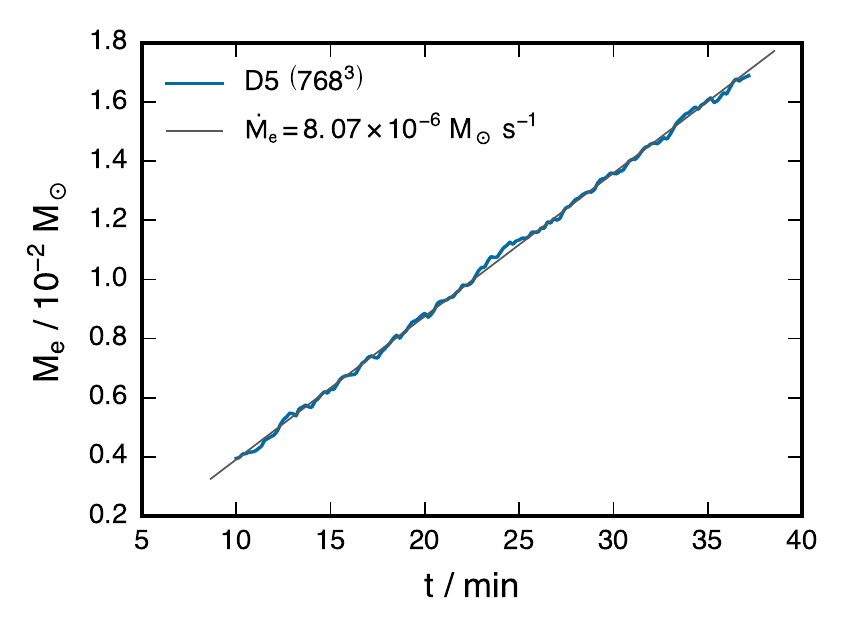}
 \includegraphics[width=0.49\linewidth, clip=true, trim = 0mm 8mm 0mm
0mm]{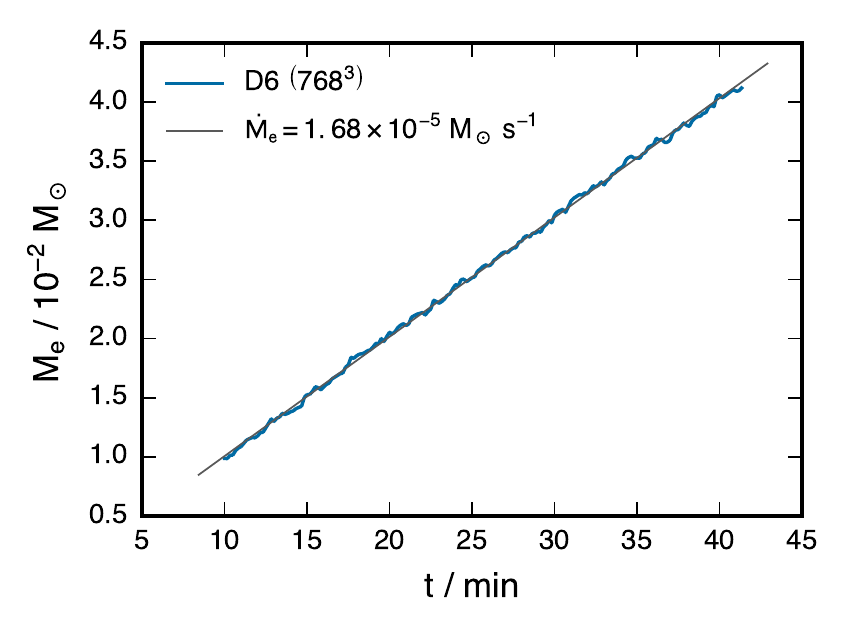}
 \includegraphics[width=0.49\linewidth, clip=true, trim = 0mm 8mm 0mm
0mm]{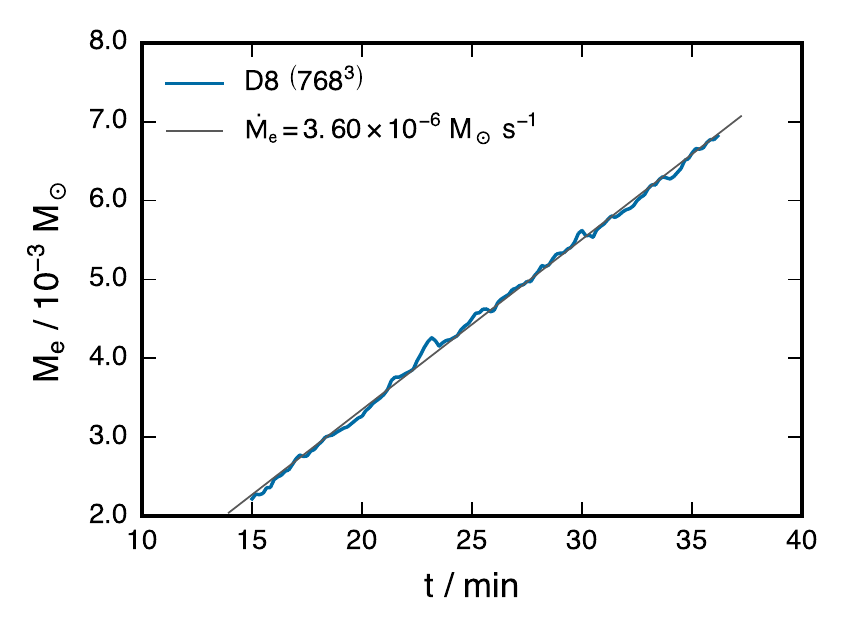}
 \includegraphics[width=0.49\linewidth, clip=true, trim = 0mm 8mm 0mm
0mm]{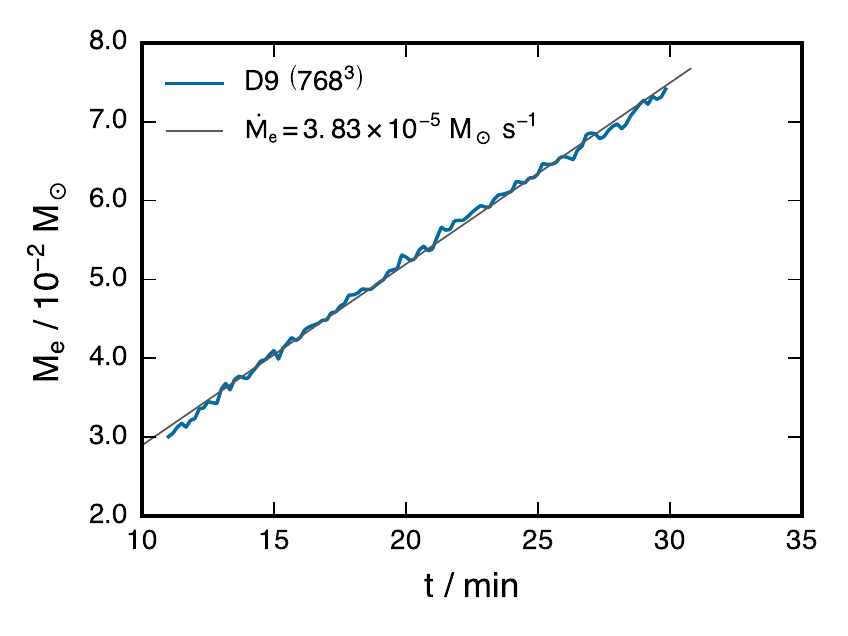}
 \includegraphics[width=0.49\linewidth]{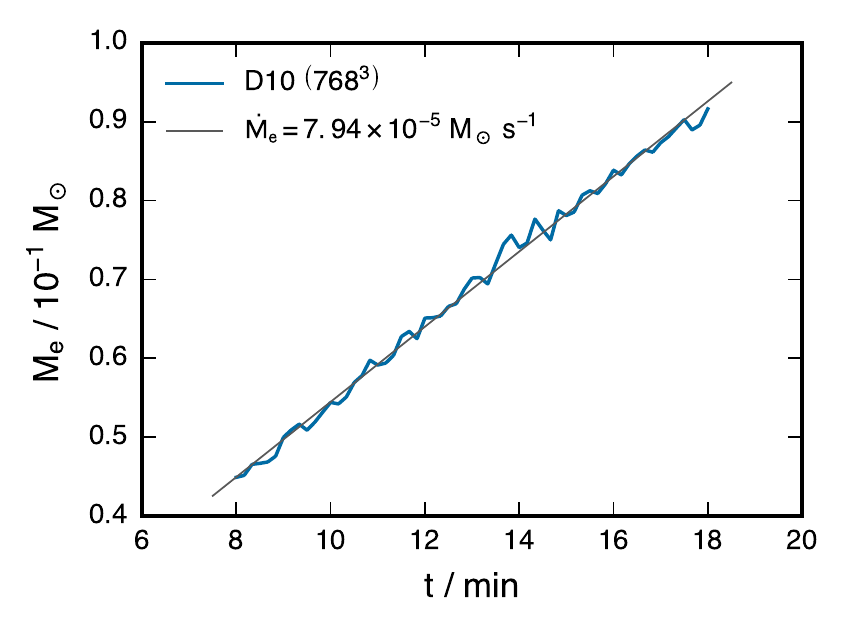}

 \caption{Entrained mass as a function of time in the runs \textsc{d5},
\textsc{d6}, \textsc{d8}, \textsc{d9}, and \textsc{d10}. The entrained mass is
fit with a straight line, whose gradient gives the entrainment rate at the
upper convective boundary of the simulation. The properties of the simulations
are given in Table~\ref{tab:run-info}}

 \label{fig:entrainment-rates-other-runs}
\end{figure*}

Table~\ref{tab:run-info} gives entrainment rates for several simulations that
have been performed as part of this work. In Fig.~\ref{fig:entrainment-rates}
the entrained mass is plotted as a function of time for the {\sc d1} and {\sc
d2} simulations with a linear fit whose gradient is the entrainment rate.
Fig.~\ref{fig:entrainment-rates-other-runs} shows the same kind of plot for the
other runs that are listed in Table~\ref{tab:run-info}.

\section{Determination of convective boundary mixing scale height}
\label{sec:f-determination}

\begin{figure}
\centering
\includegraphics[clip=true, trim=0mm 9mm 0mm 0mm]{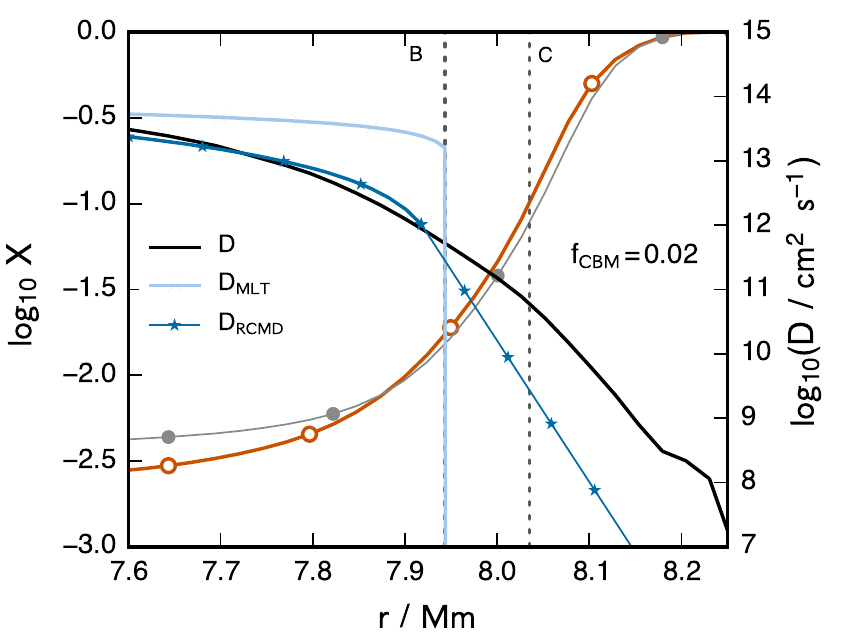}
\\
\includegraphics[clip=true, trim=0mm 9mm 0mm 0mm]{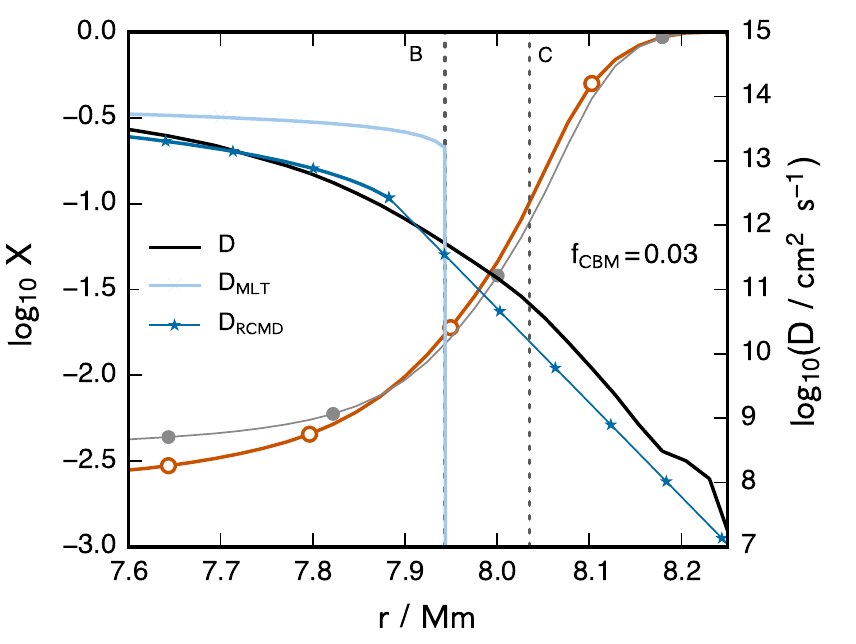}
\\
\includegraphics[]{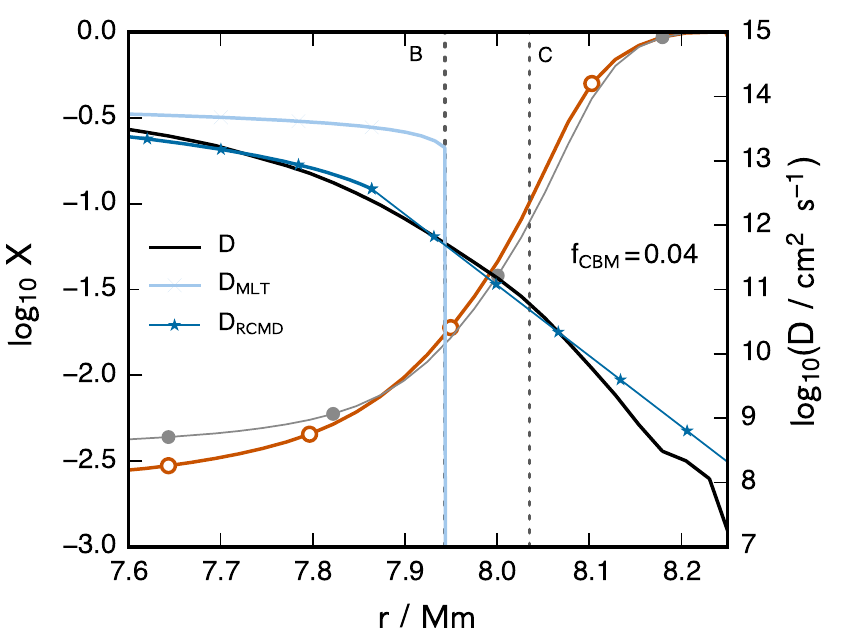}

\caption{As Fig.~\ref{fig:DDDDD_boundary}. Comparison of the diffusion
coefficient profile that reproduces the mixing of spherically-averaged 3D
hydrodynamic simulation data from the {\sc PPMstar} {\sc d1} simulation ($D$,
black solid line) with the exponentially decaying convective boundary mixing
model of \citet{freytag:96} with different CBM parameters.  While using
$f_\mathrm{CBM}=0.03$ results in slight underestimation of the diffusion
coefficient, it gives the best match to the gradient of $D$ above 8~Mm.}

\label{fig:d1_Dchoice}
\end{figure}

In Section~\ref{sec:cbm} it is described how a diffusion coefficient can be
derived that represents the 1D spherically-averaged mixing properties of the 3D
{\sc PPMstar} hydrodynamic simulations. It is shown in
Fig.~\ref{fig:DDDDD_boundary} that combining a restricted mixing length with an
exponential decay of the diffusion coefficient ($D_\mathrm{RCMD}$, solid blue
starred line) with CBM parameter $f_\mathrm{CBM}=0.03$ reproduces the
spherically-averaged mixing properties very well indeed. The value
$f_\mathrm{CBM}=0.03$ is not the result of a fitting procedure, however to the
eye it is the best fit to one decimal place. This is illustrated in
Fig.~\ref{fig:d1_Dchoice}, which shows the same plot as
Fig.~\ref{fig:DDDDD_boundary} for the {\sc d1} simulation with different values
of $f_\mathrm{CBM}$.

\clearpage
\label{lastpage}
\end{document}